# Effective Strong Dimension, Algorithmic Information, and Computational Complexity

Krishna B. Athreya [1]     John M. Hitchcock [2]     Jack H. Lutz [3]
Elvira Mayordomo [4]

[1]School of Operations Research and Industrial Engineering, Cornell University, Ithaca, NY 14853, USA and Departments of Mathematics and Statistics, Iowa State University, Ames, IA 50011, USA. kba@iastate.edu. This research was supported in part by Air Force Office of Scientific Research Grant ITSI F 49620-01-1-0076.

[2]Department of Computer Science, Iowa State University, Ames, IA 50011, USA. jhitchco@cs.iastate.edu. This research was supported in part by National Science Foundation Grant 9988483.

[3]Department of Computer Science, Iowa State University, Ames, IA 50011, USA. lutz@cs.iastate.edu. This research was supported in part by National Science Foundation Grant 9988483.

[4]Departamento de Informática e Ingeniería de Sistemas, Universidad de Zaragoza, 50015 Zaragoza, SPAIN. elvira@posta.unizar.es. This research was supported in part by Spanish Government MEC project PB98-0937-C04-02 and by National Science Foundation Grant 9988483. It was done while visiting Iowa State University.

**Abstract**

The two most important notions of fractal dimension are *Hausdorff dimension*, developed by Hausdorff (1919), and *packing dimension*, developed independently by Tricot (1982) and Sullivan (1984). Both dimensions have the mathematical advantage of being defined from measures, and both have yielded extensive applications in fractal geometry and dynamical systems.

Lutz (2000) has recently proven a simple characterization of Hausdorff dimension in terms of *gales*, which are betting strategies that generalize martingales. Imposing various computability and complexity constraints on these gales produces a spectrum of effective versions of Hausdorff dimension, including constructive, computable, polynomial-space, polynomial-time, and finite-state dimensions. Work by several investigators has already used these effective dimensions to shed significant new light on a variety of topics in theoretical computer science.

In this paper we show that packing dimension can also be characterized in terms of gales. Moreover, even though the usual definition of packing dimension is considerably more complex than that of Hausdorff dimension, our gale characterization of packing dimension is an exact dual of – and every bit as simple as – the gale characterization of Hausdorff dimension.

Effectivizing our gale characterization of packing dimension produces a variety of *effective strong dimensions*, which are exact duals of the effective dimensions mentioned above. In general (and in analogy with the classical fractal dimensions), the effective strong dimension of a set or sequence is at least as great as its effective dimension, with equality for sets or sequences that are sufficiently regular.

We develop the basic properties of effective strong dimensions and prove a number of results relating them to fundamental aspects of randomness, Kolmogorov complexity, prediction, Boolean circuit-size complexity, polynomial-time degrees, and data compression. Aside from the above characterization of packing dimension, our two main theorems are the following.

1. If $\vec{\beta} = (\beta_0, \beta_1, \ldots)$ is a computable sequence of biases that are bounded away from 0 and $R$ is random with respect to $\vec{\beta}$, then the dimension and strong dimension of $R$ are the lower and upper average entropies, respectively, of $\vec{\beta}$.

2. For each pair of $\Delta_2^0$-computable real numbers $0 \leq \alpha \leq \beta \leq 1$, there exists $A \in \text{E}$ such that the polynomial-time many-one degree of $A$ has dimension $\alpha$ in E and strong dimension $\beta$ in E.

Our proofs of these theorems use a new large deviation theorem for self-information with respect to a bias sequence $\vec{\beta}$.

# 1 Introduction

Hausdorff dimension – a powerful tool of fractal geometry developed by Hausdorff [12] in 1919 – was effectivized in 2000 by Lutz [20, 21]. This has led to a spectrum of effective versions of Hausdorff dimension, including constructive, computable, polynomial-space, polynomial-time, and finite-state dimensions. Work by several investigators has already used these effective dimensions to illuminate a variety of topics in algorithmic information theory and computational complexity [20, 21, 1, 7, 26, 16, 15, 11, 13, 14, 10]. (See [25] for a survey of some of these results.) This work has also underscored and renewed the importance of earlier work by Ryabko [27, 28, 29, 30], Staiger [36, 37, 38], and Cai and Hartmanis [5] relating Kolmogorov complexity to classical Hausdorff dimension. (See Section 6 of [21] for a discussion of this work.)

The key to all these effective dimensions is a simple characterization of classical Hausdorff dimension in terms of *gales*, which are betting strategies that generalize martingales. (Martingales, introduced by Lévy [18] and Ville [44] have been used extensively by Schnorr [31, 32, 33] and others in the investigation of randomness and by Lutz [22, 23] and others in the development of resource-bounded measure.) Given this characterization, it is a simple matter to impose computability and complexity constraints on the gales to produce the above-mentioned spectrum of effective dimensions.

In the 1980s, a new concept of fractal dimension, called the packing dimension, was introduced independently by Tricot [41] and Sullivan [39]. Packing dimension shares with Hausdorff dimension the mathematical advantage of being based on a measure. Over the past two decades, despite its greater complexity (requiring an extra optimization over all countable decompositions of a set in its definition), packing dimension has become, next to Hausdorff dimension, the most important notion of fractal dimension, yielding extensive applications in fractal geometry and dynamical systems [8, 9].

The main result of this paper is a proof that packing dimension can also be characterized in terms of gales. Moreover, notwithstanding the greater complexity of packing dimension's definition (and the greater complexity of its behavior on compact sets, as established by Mattila and Mauldin [24]), our gale characterization of packing dimension is an exact dual of – and every bit as simple as – the gale characterization of Hausdorff dimension. (This duality and simplicity are in the *statement* of our gale characterization; its proof is perforce more involved than its counterpart for Hausdorff dimension.)

Effectivizing our gale characterization of packing dimension produces for each of the effective dimensions above an *effective strong dimension* that is its exact dual. Just as the Hausdorff dimension of a set is bounded above by its packing dimension, the effective dimension of a set is bounded above by its effective strong dimension. Moreover, just as in the classical case, the effective dimension coincides with the strong effective dimension for sets that are sufficiently regular.

After proving our gale characterization and developing the effective strong dimensions and some of their basic properties, we prove a number of results relating them to fundamental aspects of randomness, Kolmogorov complexity, prediction, Boolean circuit-size complexity, polynomial-time degrees, and data compression. Our two main theorems along these lines are the following.

1. If $\delta > 0$ and $\vec{\beta} = (\beta_0, \beta_1, \ldots)$ is a computable sequence of biases with each $\beta_i \in [\delta, \frac{1}{2}]$, then every sequence $R$ that is random with respect to $\vec{\beta}$ has dimension
$$\dim(R) = \liminf_{n \to \infty} \frac{1}{n} \sum_{i=0}^{n-1} \mathcal{H}(\beta_i)$$

and strong dimension

$$\text{Dim}(R) = \limsup_{n \to \infty} \frac{1}{n} \sum_{i=0}^{n-1} \mathcal{H}(\beta_i),$$

where $\mathcal{H}(\beta_i)$ is the Shannon entropy of $\beta_i$.

2. For every pair of $\Delta_2^0$-computable real numbers $0 \leq \alpha \leq \beta \leq 1$ there is a decision problem $A \in \text{E}$ such that the polynomial-time many-one degree of $A$ has dimension $\alpha$ in E and strong dimension $\beta$ in E.

In order to prove these theorems, we prove a new large deviation theorem for the self-information $\log \frac{1}{\mu^{\vec{\beta}}(w)}$, where $\vec{\beta}$ is as in 1 above.

A corollary of theorem 1 above is that, if the average entropies $\frac{1}{n} \sum_{i=0}^{n-1} \mathcal{H}(\beta_i)$ converge to a limit $\overline{H}(\vec{\beta})$ as $n \to \infty$, then $\dim(R) = \text{Dim}(R) = \overline{H}(\vec{\beta})$. Since the convergence of these average entropies is a much weaker condition than the convergence of the biases $\beta_n$ as $n \to \infty$, this corollary substantially strengthens Theorem 7.7 of [21].

Our remaining results are much easier to prove, but their breadth makes a strong *prima facie* case for the utility of effective strong dimension. They in several cases explain dual concepts that had been curiously neglected in earlier work, and they are likely to be useful in future applications. It is to be hoped that we are on the verge of seeing the full force of fractal geometry applied fruitfully to difficult problems in the theory of computing.

## 2 Preliminaries

We use the set $\mathbb{Z}$ of integers, the set $\mathbb{Z}^+$ of (strictly) positive integers, the set $\mathbb{N}$ of natural numbers (i.e., nonnegative integers), the set $\mathbb{Q}$ of rational numbers, the set $\mathbb{R}$ of real numbers, and the set $[0, \infty)$ of nonnegative reals. All logarithms in this paper are base 2.

A *string* is a finite, binary string $w \in \{0,1\}^*$. We write $|w|$ for the length of a string $w$ and $\lambda$ for the empty string. For $i, j \in \{0, \ldots, |w| - 1\}$, we write $w[i..j]$ for the string consisting of the $i^{\text{th}}$ through the $j^{\text{th}}$ bits of $w$ and $w[i]$ for $w[i..i]$, the $i^{\text{th}}$ bit of $w$. Note that the $0^{\text{th}}$ bit $w[0]$ is the leftmost bit of $w$ and that $w[i..j] = \lambda$ if $i > j$. A *sequence* is an infinite, binary sequence. If $S$ is a sequence and $i, j \in \mathbb{N}$, then the notations $S[i..j]$ and $S[i]$ are defined exactly as for strings. We work in the *Cantor space* $\mathbf{C}$ consisting of all sequences. A string $w \in \{0,1\}^*$ is a *prefix* of a sequence $S \in \mathbf{C}$, and we write $w \sqsubseteq S$, if $S[0..|w| - 1] = w$. The *cylinder generated by* a string $w \in \{0,1\}^*$ is $\mathbf{C}_w = \{S \in \mathbf{C} | w \sqsubseteq S\}$. Note that $\mathbf{C}_\lambda = \mathbf{C}$.

Given a set $A \subseteq \{0,1\}^*$ and $n \in \mathbb{N}$, we use the abbreviations $A_{=n} = A \cap \{0,1\}^n$ and $A_{\leq n} = A \cap \{0,1\}^{\leq n}$. A *prefix set* is a set $A \subseteq \{0,1\}^*$ such that no element of $A$ is a prefix of another element of $A$.

For each $i \in \mathbb{N}$ we define a class $G_i$ of functions from $\mathbb{N}$ into $\mathbb{N}$ as follows.

$$\begin{aligned} G_0 &= \{f \mid (\exists k)(\forall^\infty n) f(n) \leq kn\} \\ G_{i+1} &= 2^{G_i(\log n)} = \{f \mid (\exists g \in G_i)(\forall^\infty n) f(n) \leq 2^{g(\log n)}\} \end{aligned}$$

We also define the functions $\hat{g}_i \in G_i$ by $\hat{g}_0(n) = 2n$, $\hat{g}_{i+1}(n) = 2^{\hat{g}_i(\log n)}$. We regard the functions in these classes as growth rates. In particular, $G_0$ contains the linearly bounded growth rates and $G_1$ contains the polynomially bounded growth rates. It is easy to show that each $G_i$ is closed under composition, that each $f \in G_i$ is $o(\hat{g}_{i+1})$, and that each $\hat{g}_i$ is $o(2^n)$. Thus $G_i$ contains superpolynomial growth rates for all $i > 1$, but all growth rates in the $G_i$-hierarchy are subexponential.

Let CE be the class of computably enumerable languages. Within the class DEC of all decidable languages, we are interested in the exponential complexity classes $E_i = \text{DTIME}(2^{G_{i-1}})$ and $E_i\text{SPACE} = \text{DSPACE}(2^{G_{i-1}})$ for $i \geq 1$. The much-studied classes $E = E_1 = \text{DTIME}(2^{\text{linear}})$, $E_2 = \text{DTIME}(2^{\text{polynomial}})$, and $\text{ESPACE} = E_1\text{SPACE} = \text{DSPACE}(2^{\text{linear}})$ are of particular interest.

We use the following classes of functions.

$$\begin{aligned}
\text{all} &= \{f \mid f : \{0,1\}^* \to \{0,1\}^*\} \\
\text{comp} &= \{f \in \text{all} \mid f \text{ is computable}\} \\
\text{p}_i &= \{f \in \text{all} \mid f \text{ is computable in } G_i \text{ time}\} \ (i \geq 1) \\
\text{p}_i\text{space} &= \{f \in \text{all} \mid f \text{ is computable in } G_i \text{ space}\} \ (i \geq 1)
\end{aligned}$$

(The length of the output *is* included as part of the space used in computing $f$.) We write p for $\text{p}_1$ and pspace for $\text{p}_1\text{space}$.

A *constructor* is a function $\delta : \{0,1\}^* \to \{0,1\}^*$ that satisfies $x \sqsubsetneq \delta(x)$ for all $x$. The *result* of a constructor $\delta$ (i.e., the language *constructed* by $\delta$) is the unique language $R(\delta)$ such that $\delta^n(\lambda) \sqsubseteq R(\delta)$ for all $n \in \mathbb{N}$. Intuitively, $\delta$ constructs $R(\delta)$ by starting with $\lambda$ and then iteratively generating successively longer prefixes of $R(\delta)$. We write $R(\Delta)$ for the set of languages $R(\delta)$ such that $\delta$ is a constructor in $\Delta$. The following facts are the reason for our interest in the above-defined classes of functions.

$R(\text{all}) = \mathbf{C}$.
$R(\text{comp}) = \text{DEC}$.
For $i \geq 1$, $R(\text{p}_i) = E_i$.
For $i \geq 1$, $R(\text{p}_i\text{space}) = E_i\text{SPACE}$.

If $D$ is a discrete domain (such as $\mathbb{N}, \{0,1\}^*, \mathbb{N} \times \{0,1\}^*$, etc.), then a function $f : D \longrightarrow [0, \infty)$ is $\Delta$-*computable* if there is a function $\hat{f} : \mathbb{N} \times D \longrightarrow \mathbb{Q} \cap [0, \infty)$ such that $|\hat{f}(r, x) - f(x)| \leq 2^{-r}$ for all $r \in \mathbb{N}$ and $x \in D$ and $\hat{f} \in \Delta$ (with $r$ coded in unary and the output coded in binary). We say that $f$ is *exactly* $\Delta$-*computable* if $f : D \longrightarrow \mathbb{Q} \cap [0, \infty)$ and $f \in \Delta$. We say that $f$ is *lower semicomputable* if there is a computable function $\hat{f} : D \times \mathbb{N} \to \mathbb{Q}$ such that

(a) for all $(x, t) \in D \times \mathbb{N}$, $\hat{f}(x, t) \leq \hat{f}(x, t+1) < f(x)$, and

(b) for all $x \in D$, $\lim_{t \to \infty} \hat{f}(x, t) = f(x)$.

Let $k$ be a positive integer. A *k-account finite-state gambler (k-account FSG)* is a tuple $G = (Q, \delta, \beta, q_0, \vec{c_0})$ where

- $Q$ is a nonempty, finite set of states,
- $\delta : Q \times \{0,1\} \to Q$ is the transition function,
- $\beta : \{1, \ldots, k\} \times Q \times \{0,1\} \to \mathbb{Q} \cap [0,1]$ is the betting function,
- $q_0 \in Q$ is the initial state, and
- $\vec{c_0}$ is the initial capital vector, a sequence of $k$ nonnegative rational numbers.

The betting function satisfies $\beta(i, q, 0) + \beta(i, q, 1) = 1$ for each $q \in Q$ and $1 \leq i \leq k$. We use the standard extension $\delta^* : \Sigma^* \to Q$ of $\delta$ defined recursively by $\delta^*(\lambda) = q_0$ and $\delta^*(wb) = \delta(\delta^*(w), b)$ for all $w \in \{0,1\}^*$ and $b \in \{0,1\}$.

# 3 Fractal Dimensions

In this section we briefly review the classical definitions of some fractal dimensions and the relationships among them. Since we are primarily interested in binary sequences and (equivalently) decision problems, we focus on fractal dimension in the Cantor space $\mathbf{C}$.

For each $k \in \mathbb{N}$, we let $\mathcal{A}_k$ be the collection of all prefix sets $A$ such that $A_{<k} = \emptyset$. For each $X \subseteq \mathbf{C}$, we then define the families

$$\mathcal{A}_k(X) = \left\{ A \in \mathcal{A}_k \,\bigg|\, X \subseteq \bigcup_{w \in A} \mathbf{C}_w \right\},$$

$$\mathcal{B}_k(X) = \left\{ A \in \mathcal{A}_k \,|\, (\forall w \in A) \mathbf{C}_w \cap X \neq \emptyset \right\}.$$

If $A \in \mathcal{A}_k(X)$, then we say that the prefix set $A$ *covers* the set $X$. If $A \in \mathcal{B}_k(X)$, then we call the prefix set $A$ a *packing* of $X$. For $X \in \mathbf{C}$, $s \in [0, \infty)$, and $k \in \mathbb{N}$, we then define

$$H_k^s(X) = \inf_{A \in \mathcal{A}_k(X)} \sum_{w \in A} 2^{-s|w|},$$

$$P_k^s(X) = \sup_{A \in \mathcal{B}_k(X)} \sum_{w \in A} 2^{-s|w|}.$$

Since $H_k^s(X)$ and $P_k^s(X)$ are monotone in $k$, the limits

$$H^s(X) = \lim_{k \to \infty} H_k^s(X),$$

$$P_\infty^s(X) = \lim_{k \to \infty} P_k^s(X)$$

exist, though they may be infinite. We then define

$$P^s(X) = \inf \left\{ \sum_{i=0}^{\infty} P_\infty^s(X_i) \,\bigg|\, X \subseteq \bigcup_{i=0}^{\infty} X_i \right\}. \tag{3.1}$$

The set functions $H^s$ and $P^s$ have the technical properties of an outer measure [8], and the (possibly infinite) quantities $H^s(X)$ and $P^s(X)$ are thus known as the *s-dimensional Hausdorff (outer) measure* of $X$ and the *s-dimensional packing (outer) measure* of $X$, respectively. The set function $P_\infty^s$ is *not* an outer measure; this is the reason for the extra optimization (3.1) in the definition of the packing measure.

**Definition.** *Let $X \subseteq \mathbf{C}$.*

1. *The* Hausdorff dimension *of $X$ is $\dim_{\mathrm{H}}(X) = \inf\{s \in [0, \infty) | H^s(X) = 0\}$.*

2. *The* packing dimension *of $X$ is $\dim_{\mathrm{P}}(X) = \inf\{s \in [0, \infty) | P^s(X) = 0\}$.*

The proof of our main result uses a well-known characterization of packing dimension as a modified box dimension. For each $X \subseteq \mathbf{C}$ and $n \in \mathbb{N}$, let

$$N_n(X) = \left| \{ w \in \{0,1\}^n | (\exists S \in X) w \sqsubseteq S \} \right|.$$

Then the *upper box dimension* of $X$ is

$$\overline{\dim}_{\mathrm{B}}(X) = \limsup_{n \to \infty} \frac{\log N_n(X)}{n}. \tag{3.2}$$

The lower box dimension $\underline{\dim}_{\mathrm{B}}(X)$, which we do not use here, is obtained by using a limit inferior in place of the limit superior in (3.2). When $\underline{\dim}_{\mathrm{B}}(X) = \overline{\dim}_{\mathrm{B}}(X)$, this quantity, written $\dim_{\mathrm{B}}(X)$, is called the box dimension of $X$.

Box dimensions are over 60 years old, have been re-invented many times, and have been named many things, including Minkowski dimension, Kolmogorov entropy, Kolmogorov dimension, topological entropy, metric dimension, logarithmic density, and information dimension. Box dimensions are often used in practical applications of fractal geometry because they are easy to estimate, but they are not well-behaved mathematically. The *modified upper box dimension*

$$\overline{\dim}_{\mathrm{MB}}(X) = \inf \left\{ \sup_i \overline{\dim}_{\mathrm{B}}(X_i) \,\Big|\, X \subseteq \bigcup_{i=0}^{\infty} X_i \right\} \tag{3.3}$$

is much better behaved. (Note that (3.3), like (3.1), is an optimization over all countable decompositions of $X$.) In fact, the following relations are well-known [8].

**Theorem 3.1.** *For all $X \subseteq \mathbf{C}$, $0 \leq \dim_{\mathrm{H}}(X) \leq \overline{\dim}_{\mathrm{MB}}(X) = \dim_{\mathrm{P}}(X) \leq \overline{\dim}_{\mathrm{B}}(X) \leq 1$.*

The above dimensions are *monotone*, i.e., $X \subseteq Y$ implies $\dim(X) \leq \dim(Y)$, and *stable*, i.e., $\dim(X \cup Y) = \max\{\dim(X), \dim(Y)\}$. The Hausdorff and packing dimensions are also *countably stable*, i.e., $\dim(\cup_{i=0}^{\infty} X_i) = \sup\{\dim(X_i) | i \in \mathbb{N}\}$.

## 4 Gale Characterizations

In this section we review the gale characterization of Hausdorff dimension and prove our main theorem, which is the dual gale characterization of packing dimension.

**Definition.** *Let $s \in [0, \infty)$.*

1. *An $s$-supergale is a function $d : \{0,1\}^* \longrightarrow [0, \infty)$ that satisfies the condition*

$$d(w) \geq 2^{-s}[d(w0) + d(w1)] \tag{4.1}$$

   *for all $w \in \{0,1\}^*$.*

2. *An $s$-gale is an $s$-supergale that satisfies (4.1) with equality for all $w \in \{0,1\}^*$.*

3. *A supermartingale is a 1-supergale.*

4. *A martingale is a 1-gale.*

Intuitively, we regard a supergale $d$ as a strategy for betting on the successive bits of a sequence $S \in \mathbf{C}$. More specifically $d(w)$ is the amount of capital that $d$ has after betting on the prefix $w$ of $S$. If $s = 1$, then the right-hand side of (4.1) is the conditional expectation of $d(wb)$ given that $w$ has occurred (when $b$ is a uniformly distributed binary random variable). Thus a martingale models a gambler's capital when the payoffs are fair. (The expected capital after the bet is the actual capital before the bet.) In the case of an $s$-gale, if $s < 1$, the payoffs are less than fair; if $s > 1$, the payoffs are more than fair.

We use the following known generalization of the Kraft inequality.

**Lemma 4.1.** (Lutz [20]) *Let $s \in [0, \infty)$. If $d$ is an $s$-supergale and $B \subseteq \{0,1\}^*$ is a prefix set, then for all $w \in \{0,1\}^*$, $\sum_{u \in B} 2^{-s|u|} d(wu) \leq d(w)$.*

We now define two criteria for the success of a gale or supergale.

**Definition.** *Let $d$ be an $s$-supergale, where $s \in [0, \infty)$.*

1. *We say that $d$ succeeds on a sequence $S \in \mathbf{C}$ if*

$$\limsup_{n \to \infty} d(S[0..n-1]) = \infty. \quad (4.2)$$

*The success set of $d$ is $S^\infty[d] = \{S \in \mathbf{C} | d \text{ succeeds on } S\}$.*

2. *We say that $d$ succeeds strongly on a sequence $S \in \mathbf{C}$ if*

$$\liminf_{n \to \infty} d(S[0..n-1]) = \infty. \quad (4.3)$$

*The strong success set of $d$ is $S^\infty_{\text{str}}[d] = \{S \in \mathbf{C} | d \text{ succeeds strongly on } S\}$.*

We have written conditions (4.2) and (4.3) in a fashion that emphasizes their duality. Condition (4.2) says simply that the set of values $d(S[0..n-1])$ is unbounded, while condition (4.3) says that $d(S[0..n-1]) \to \infty$ as $n \to \infty$.

**Notation.** Let $X \subseteq \mathbf{C}$.

1. $\mathcal{G}(X)$ is the set of all $s \in [0, \infty)$ for which there exists an $s$-gale $d$ such that $X \subseteq S^\infty[d]$.

2. $\mathcal{G}^{\text{str}}(X)$ is the set of all $s \in [0, \infty)$ for which there exists an $s$-gale $d$ such that $X \subseteq S^\infty_{\text{str}}[d]$.

3. $\widehat{\mathcal{G}}(X)$ is the set of all $s \in [0, \infty)$ for which there exists an $s$-supergale $d$ such that $X \subseteq S^\infty[d]$.

4. $\widehat{\mathcal{G}}^{\text{str}}(X)$ is the set of all $s \in [0, \infty)$ for which there exists an $s$-supergale $d$ such that $X \subseteq S^\infty_{\text{str}}[d]$.

Note that $s' \geq s \in \mathcal{G}(X)$ implies that $s' \in \mathcal{G}(X)$, and similarly for the classes $\mathcal{G}^{\text{str}}(X)$, $\widehat{\mathcal{G}}(X)$, and $\widehat{\mathcal{G}}^{\text{str}}(X)$. The following fact is also clear.

**Observation 4.2.** *For all $X \subseteq \mathbf{C}$, $\mathcal{G}(X) = \widehat{\mathcal{G}}(X)$ and $\mathcal{G}^{\text{str}}(X) = \widehat{\mathcal{G}}^{\text{str}}(X)$.*

For Hausdorff dimension, we have the following known fact.

**Theorem 4.3.** *(Gale Characterization of Hausdorff Dimension – Lutz [20]) For all $X \subseteq \mathbf{C}$, $\dim_{\text{H}}(X) = \inf \mathcal{G}(X)$.*

Our main result is the following dual of Theorem 4.3.

**Theorem 4.4.** *(Gale Characterization of Packing Dimension) For all $X \subseteq \mathbf{C}$, $\dim_{\text{P}}(X) = \inf \mathcal{G}^{\text{str}}(X)$.*

By Observation 4.2, we could equivalently use $\widehat{\mathcal{G}}(X)$ and $\widehat{\mathcal{G}}^{\text{str}}(X)$ in Theorems 4.3 and 4.4, respectively. We will use the following lemma to prove Theorem 4.4.

**Lemma 4.5.** *For each family of sets $\{X_k \subseteq \mathbf{C} | k \in \mathbb{N}\}$, $\inf \mathcal{G}^{\text{str}} \left( \bigcup_k X_k \right) = \sup_k \inf \mathcal{G}^{\text{str}}(X_k)$.*

*Proof.* The inequality $\inf \mathcal{G}^{\mathrm{str}}(\bigcup_k X_k) \geq \sup_k \inf \mathcal{G}^{\mathrm{str}}(X_k)$ holds trivially.

To prove that $\inf \mathcal{G}^{\mathrm{str}}(\bigcup_k X_k) \leq \sup_k \inf \mathcal{G}^{\mathrm{str}}(X_k)$, let $s > \sup_k \inf \mathcal{G}^{\mathrm{str}}(X_k)$. Then for each $k \in \mathbb{N}$ there is an $s$-gale $d_k$ such that $X_k \subseteq S_{\mathrm{str}}^\infty[d_k]$. We define an $s$-gale $d$ by

$$d(w) = \sum_{k \in \mathbb{N}} \frac{2^{-k}}{d_k(\lambda)} \cdot d_k(w)$$

for all $w \in \{0,1\}^*$. Then for each $k$, for any $S \in X_k$, we have

$$d(S[0..n-1]) \geq \frac{2^{-k}}{d_k(\lambda)} \cdot d_k(S[0..n-1])$$

for all $n$, so $S \in S_{\mathrm{str}}^\infty[d]$. Therefore $\bigcup_k X_k \subseteq S_{\mathrm{str}}^\infty[d]$ and the lemma follows. $\square$

*Proof of Theorem 4.4.* Let $X \subseteq \mathbf{C}$. By Theorem 3.1, it suffices to show that $\overline{\dim}_{\mathrm{MB}}(X) = \inf \mathcal{G}^{\mathrm{str}}(X)$.

To see that $\overline{\dim}_{\mathrm{MB}}(X) \leq \inf \mathcal{G}^{\mathrm{str}}(X)$, let $s > \inf \mathcal{G}^{\mathrm{str}}(X)$. It suffices to show that $\overline{\dim}_{\mathrm{MB}}(X) \leq s$. By our choice of $s$, there is an $s$-gale $d$ such that $X \subseteq S_{\mathrm{str}}^\infty[d]$. For each $n \in \mathbb{N}$, let

$$B_n = \{w \in \{0,1\}^n \mid d(w) > d(\lambda)\}$$

and

$$Y_n = \{S \in \mathbf{C} \mid S[0..n-1] \in B_n\}.$$

For each $i \in \mathbb{N}$, let

$$X_i = \bigcap_{n=i}^\infty Y_n,$$

and note that

$$X \subseteq \bigcup_{i=0}^\infty X_i. \tag{4.4}$$

For all $n \geq i \in \mathbb{N}$, we have $X_i \subseteq Y_n$, whence the generalized Kraft inequality (Lemma 4.1) tells us that

$$N_n(X_i) \leq N_n(Y_n) = |B_n| < 2^{sn}.$$

It follows that, for all $i \in \mathbb{N}$,

$$\overline{\dim}_{\mathrm{B}}(X_i) = \limsup_{n \to \infty} \frac{\log N_n(X_i)}{n} \leq s,$$

whence by (4.4),

$$\overline{\dim}_{\mathrm{MB}}(X) \leq \sup_{i \in \mathbb{N}} \overline{\dim}_{\mathrm{B}}(X_i) \leq s.$$

To see that $\inf \mathcal{G}^{\mathrm{str}}(X) \leq \overline{\dim}_{\mathrm{MB}}(X)$, let $s > s' > s'' > \overline{\dim}_{\mathrm{MB}}(X)$. It suffices to show that $\inf \mathcal{G}^{\mathrm{str}}(X) \leq s$. Since $s'' > \overline{\dim}_{\mathrm{MB}}(X)$, there exist sets $X_0, X_1, \ldots \subseteq \mathbf{C}$ such that $X = \bigcup_{i=0}^\infty X_i$ and $\overline{\dim}_{\mathrm{B}}(X_i) < s''$ for all $i \in \mathbb{N}$. By Lemma 4.5, it suffices to show that $s \in \mathcal{G}^{\mathrm{str}}(X_i)$ for all $i \in \mathbb{N}$.

Fix $i \in \mathbb{N}$. Since $\overline{\dim}_{\mathrm{B}}(X_i) < s''$, there exists $n_0 \in \mathbb{N}$ such that, for all $n \geq n_0$, $\frac{\log N_n(X_i)}{n} < s''$, i.e., $N_n(X_i) < 2^{s''n}$. For each $n \geq n_0$, let

$$A_n = \{S[0..n-1] \mid S \in X_i\}$$

(noting that $|A_n| = N_n(X_i)$), and define $d_n : \{0,1\}^* \to [0, \infty)$ by

$$d_n(w) = \begin{cases} 2^{(s-s')|w|} \sum_{\substack{u \\ wu \in A_n}} 2^{-s'|u|} & \text{if } |w| \leq n \\ 2^{(s-1)(|w|-n)} d_n(w[0..n-1]) & \text{if } |w| > n. \end{cases}$$

It is routine to verify that $d_n$ is an $s$-gale for each $n \geq n_0$. Note also that $d_n(w) = 2^{(s-s')n}$ for all $n \geq n_0$ and $w \in A_n$. Let $d = \sum_{n=n_0}^{\infty} d_n$. Then

$$\begin{aligned} d(\lambda) &= \sum_{n=n_0}^{\infty} d_n(\lambda) = \sum_{n=n_0}^{\infty} |A_n| 2^{-s'n} = \sum_{n=n_0}^{\infty} N_n(X_i) 2^{-s'n} \\ &< \sum_{n=n_0}^{\infty} 2^{(s''-s')n} < \infty, \end{aligned}$$

so $d$ is an $s$-gale by linearity. Let $S \in X_i$. Then, for all $n \geq n_0$, $S[0..n-1] \in A_n$, so

$$d(S[0..n-1]) \geq d_n(S[0..n-1]) \geq 2^{(s-s')n}.$$

Thus $S \in S_{\text{str}}^{\infty}[d]$. This shows that $X_i \subseteq S_{\text{str}}^{\infty}[d]$, whence $s \in \mathcal{G}^{\text{str}}(X_i)$. □

## 5 Effective Strong Dimensions

Theorem 4.3 has been used to effectivize Hausdorff dimension at a variety of levels. In this section we review these effective dimensions while using Theorem 4.4 to develop the dual effective strong dimensions.

We define a gale or supergale to be *constructive* if it is lower semicomputable. For any $s \in [0, \infty)$ and any $k$-account FSG $G$ an $s$-gale $d_G^{(s)}$ is defined as follows [7]. (Recall that finite-state gamblers were defined in Section 2.) For each $1 \leq i \leq k$ we define an $s$-gale $d_{G,i}^{(s)}$ by the recursion

$$d_{G,i}^{(s)}(\lambda) = c_{0,i}$$

$$d_{G,i}^{(s)}(wb) = 2^s d_{G,i}^{(s)}(w) \beta(i, \delta^*(w), b)$$

for all $w \in \{0,1\}^*$ and $b \in \{0,1\}$. Then

$$d_G^{(s)} = \sum_{i=1}^{k} d_{G,i}^{(s)}.$$

We define an $s$-gale $d$ to be *finite-state* if there is a finite-state gambler (FSG) $G$ such that $d_G^{(s)} = d$. For the rest of this paper, $\Delta$ denotes one of the classes all, comp, p, pspace, $p_2$, $p_2$space, etc. defined in Section 2.

For each $\Gamma \in \{\text{constr}, \Delta, \text{FS}\}$ and $X \subseteq \mathbf{C}$, we define the sets $\mathcal{G}_\Gamma(X)$, $\mathcal{G}_\Gamma^{\text{str}}(X)$, $\widehat{\mathcal{G}}_\Gamma(X)$, and $\widehat{\mathcal{G}}_\Gamma^{\text{str}}(X)$ just as the classes $\mathcal{G}(X)$, $\mathcal{G}^{\text{str}}(X)$, $\widehat{\mathcal{G}}(X)$, and $\widehat{\mathcal{G}}^{\text{str}}(X)$ were defined in Section 4, but with the following modifications.

(i) If $\Gamma = \text{constr}$, then $d$ is required to be constructive.

(ii) If $\Gamma = \Delta$, then $d$ is required to be $\Delta$-computable.

(iii) In $\mathcal{G}_{\text{FS}}(X)$ and $\mathcal{G}_{\text{FS}}^{\text{str}}(X)$, $d$ is required to be finite-state.

(iv) $\widehat{\mathcal{G}}_{\text{FS}}(X)$ and $\widehat{\mathcal{G}}_{\text{FS}}^{\text{str}}(X)$ are not defined.

The following effectivizations of Hausdorff and packing dimension are motivated by Theorems 4.3 and 4.4.

**Definition.** *Let $X \subseteq \mathbf{C}$ and $S \in \mathbf{C}$.*

1. *[21] The* constructive dimension *of $X$ is* $\text{cdim}(X) = \inf \mathcal{G}_{\text{constr}}(X)$.

2. *The* constructive strong dimension *of $X$ is* $\text{cDim}(X) = \inf \mathcal{G}_{\text{constr}}^{\text{str}}(X)$.

3. *[21] The* dimension *of $S$ is* $\dim(S) = \text{cdim}(\{S\})$.

4. *The* strong dimension *of $S$ is* $\text{Dim}(S) = \text{cDim}(\{S\})$.

5. *[20] The* $\Delta$-dimension *of $X$ is* $\dim_\Delta(X) = \inf \mathcal{G}_\Delta(X)$.

6. *The* $\Delta$-strong dimension *of $X$ is* $\text{Dim}_\Delta(X) = \inf \mathcal{G}_\Delta^{\text{str}}(X)$.

7. *[20] The* dimension *of $X$ in $R(\Delta)$ is* $\dim(X|R(\Delta)) = \dim_\Delta(X \cap R(\Delta))$.

8. *The* strong dimension *of $X$ in $R(\Delta)$ is* $\text{Dim}(X|R(\Delta)) = \text{Dim}_\Delta(X \cap R(\Delta))$.

9. *[7] The* finite-state dimension *of $X$ is* $\dim_{\text{FS}}(X) = \inf \mathcal{G}_{\text{FS}}(X)$.

10. *The* finite-state strong dimension *of $X$ is* $\text{Dim}_{\text{FS}}(X) = \inf \mathcal{G}_{\text{FS}}^{\text{str}}(X)$.

11. *[7] The* finite-state dimension *of $S$ is* $\dim_{\text{FS}}(S) = \dim_{\text{FS}}(\{S\})$.

12. *The* finite-state strong dimension *of $S$ is* $\text{Dim}_{\text{FS}}(S) = \text{Dim}_{\text{FS}}(\{S\})$.

In parts 1,2,5, and 6 of the above definition, we could equivalently use the "hatted" sets $\widehat{\mathcal{G}}_{\text{constr}}(X)$, $\widehat{\mathcal{G}}_{\text{constr}}^{\text{str}}(X)$, $\widehat{\mathcal{G}}_\Delta(X)$, and $\widehat{\mathcal{G}}_\Delta^{\text{str}}(X)$ in place of their unhatted counterparts. In the case of parts 5 and 6, this follows from Lemma 4.7 of [20]. In the case of parts 1 and 2, it follows from the main theorem in [14] (which answered an open question in [21], where $\widehat{\mathcal{G}}_{\text{constr}}(X)$ was in fact used in defining $\text{cdim}(X)$).

The polynomial-time dimensions $\dim_p(X)$ and $\text{Dim}_p(X)$ are also called the feasible dimension and the feasible strong dimension, respectively. The notation $\dim_p(X)$ for the p-dimension is all too similar to the notation $\dim_P(X)$ for the classical packing dimension, but confusion is unlikely because these dimensions typically arise in quite different contexts.

Note that the classical Hausdorff and packing dimensions can each now be written in three different ways, i.e.,
$$\dim_H(X) = \dim_{\text{all}}(X) = \dim(X|\mathbf{C})$$
and
$$\dim_P(X) = \text{Dim}_{\text{all}}(X) = \text{Dim}(X|\mathbf{C}).$$

**Observations 5.1.** *1. Each of the dimensions that we have defined is* monotone *(e.g., $X \subseteq Y$ implies $\text{cdim}(X) \leq \text{cdim}(Y)$).*

2. Each of the effective strong dimensions is bounded below by the corresponding effective dimension (e.g., $\text{cdim}(X) \leq \text{cDim}(X)$).

3. Each of the dimensions that we have defined is nonincreasing as the effectivity constraint is relaxed (e.g., $\dim_H(X) \leq \text{cdim}(X) \leq \dim_{\text{pspace}}(X) \leq \dim_{\text{FS}}(X)$).

4. Each of the dimensions that we have defined is nonnegative and assigns $\mathbf{C}$ the dimension 1.

**Lemma 5.2.** *The finite-state dimensions are* stable, *i.e., for all $X, Y \subseteq \mathbf{C}$,*

$$\dim_{\text{FS}}(X \cup Y) = \max\{\dim_{\text{FS}}(X), \dim_{\text{FS}}(Y)\}$$

*and*

$$\text{Dim}_{\text{FS}}(X \cup Y) = \max\{\text{Dim}_{\text{FS}}(X), \text{Dim}_{\text{FS}}(Y)\}.$$

*Proof.* The stability of finite-state dimension was proved in [7]. The same arguments establish stability for finite-state strong dimension. □

**Definition.** *Let $X, X_0, X_1, X_2, \ldots \subseteq \mathbf{C}$.*

1. *We say that $X$ is a $\Delta$-union of the $\Delta$-dimensioned sets $\{X_k | k \in \mathbb{N}\}$ if $X = \bigcup_{k=0}^{\infty} X_k$ and for each $s > \sup_{k \in \mathbb{N}} \dim_\Delta(X_k)$ with $2^s$ rational, there is a function $d : \mathbb{N} \times \{0,1\}^* \to [0, \infty)$ with the following three properties.*

    (i) *$d$ is $\Delta$-computable.*
    
    (ii) *For each $k \in \mathbb{N}$, if we write $d_k(w) = d(k, w)$, then the function $d_k$ is an $s$-gale.*
    
    (iii) *For each $k \in \mathbb{N}$, $X_k \subseteq S^\infty[d_k]$.*

    *Analogously, $X$ is a $\Delta$-union of the $\Delta$-strong dimensioned sets $\{X_k | k \in \mathbb{N}\}$ if there is a $d$ with the above properties that also satisfies*

    (iv) *For each $k \in \mathbb{N}$, $X_k \subseteq S^\infty_{\text{str}}[d_k]$.*

2. *We say that $X$ is a $\Delta$-union of the sets $\{X_k | k \in \mathbb{N}\}$ dimensioned in $R(\Delta)$ if $X = \bigcup_{k=0}^{\infty} X_k$ and $X \cap R(\Delta)$ is an $\Delta$-union of the $\Delta$-dimensioned sets $\{X_k \cap R(\Delta) | k \in \mathbb{N}\}$.*

    *Analogously, $X$ is $\Delta$-union of the sets $\{X_k | k \in \mathbb{N}\}$ strong dimensioned in $R(\Delta)$ if $X = \bigcup_{k=0}^{\infty} X_k$ and $X \cap R(\Delta)$ is an $\Delta$-union of the $\Delta$-strong dimensioned sets $\{X_k \cap R(\Delta) | k \in \mathbb{N}\}$.*

**Lemma 5.3.** *The dimensions defined from $\Delta$ are $\Delta$-countably stable, i.e., if $X$ is a $\Delta$-union of the $\Delta$-dimensioned sets $X_0, X_1, X_2, \ldots$, then*

$$\dim_\Delta(X) = \sup_{k \in \mathbb{N}} \dim_\Delta(X_k),$$

*and if $X$ is a $\Delta$-union of the $\Delta$-strong dimensioned sets $X_0, X_1, X_2, \ldots$, then*

$$\text{Dim}_\Delta(X) = \sup_{k \in \mathbb{N}} \text{Dim}_\Delta(X_k),$$

*and similarly for dimension and strong dimension in $R(\Delta)$.*

*Proof.* The stability of $\dim_\Delta$ over $\Delta$-unions was proved in [20]. The proof for strong dimension is analogous. □

**Lemma 5.4.** *The constructive dimensions are* absolutely stable, *i.e., for all $X \subseteq \mathbf{C}$,*

$$\operatorname{cdim}(X) = \sup_{S \in X} \dim(S)$$

*and*

$$\operatorname{cDim}(X) = \sup_{S \in X} \operatorname{Dim}(S).$$

*Proof.* The absolute stability of constructive dimension was proved in [21] using optimal constructive supergales. The same argument works for constructive strong dimension. □

## 6 Algorithmic Information

In this section we present a variety of results and observations in which constructive and computable strong dimensions illuminate or clarify various aspects of algorithmic information theory. Included is our second main theorem, which says that every sequence that is random with respect to a computable sequence of biases $\beta_i \in [\delta, 1/2]$ has the lower and upper average entropies of $(\beta_0, \beta_1, \ldots)$ as its dimension and strong dimension, respectively. We also present a result in which finite-state strong dimension clarifies an issue in data compression.

Mayordomo [26] proved that for all $S \in \mathbf{C}$,

$$\dim(S) = \liminf_{n \to \infty} \frac{K(S[0..n-1])}{n}, \tag{6.1}$$

where $K(w)$ is the Kolmogorov complexity of $w$ [19]. Subsequently, Lutz [21] used termgales to define the dimension $\dim(w)$ of each (finite!) string $w \in \{0,1\}^*$ and proved that

$$\dim(S) = \liminf_{n \to \infty} \dim(S[0..n-1]) \tag{6.2}$$

for all $S \in \mathbf{C}$ and

$$K(w) = |w|\dim(w) \pm O(1) \tag{6.3}$$

for all $w \in \{0,1\}^*$, thereby giving a second proof of (6.1). The following theorem is a dual of (6.2) that yields a dual of (6.1) as a corollary.

**Theorem 6.1.** *For all $S \in \mathbf{C}$,*

$$\operatorname{Dim}(S) = \limsup_{n \to \infty} \dim(S[0..n-1]).$$

*Proof.* This proof is analogous to the one for the dual statement (6.2) given in [21]. □

**Corollary 6.2.** *For all $S \in \mathbf{C}$,*

$$\operatorname{Dim}(S) = \limsup_{n \to \infty} \frac{K(S[0..n-1])}{n}.$$

By Corollary 6.2, the "upper algorithmic dimension" defined by Tadaki [40] is precisely the constructive strong dimension.

The rate at which a gambler can increase its capital when betting in a given situation is a fundamental concern of classical and algorithmic information and computational learning theories. In the setting of constructive gamblers, the following quantities are of particular relevance.

**Definition.** *Let $d$ be a supermartingale, let $S \in \mathbf{C}$, and let $X \subseteq \mathbf{C}$.*

1. *The* lower $d$-Lyapunov exponent *of $S$ is $\lambda_d(S) = \liminf_{n \to \infty} \frac{\log d(S[0..n-1])}{n}$.*

2. *The* upper $d$-Lyapunov exponent *of $S$ is $\Lambda_d(S) = \limsup_{n \to \infty} \frac{\log d(S[0..n-1])}{n}$.*

3. *The* lower Lyapunov exponent *of $S$ is $\lambda(S) = \sup\{\lambda_d(S) | d \text{ is a constructive supermartingale}\}$.*

4. *The* upper Lyapunov exponent *of $S$ is $\Lambda(S) = \sup\{\Lambda_d(S) | d \text{ is a constructive supermartingale}\}$.*

5. *The* lower Lyapunov exponent *of $X$ is $\lambda(X) = \inf_{S \in X} \lambda(S)$.*

6. *The* upper Lyapunov exponent *of $X$ is $\Lambda(X) = \inf_{S \in X} \Lambda(S)$.*

Lyapunov exponents such as these were investigated by Schnorr [32, 34], Ryabko [30], and Staiger [37, 38] (using slightly different notations) prior to the effectivization of Hausdorff dimension. The quantities $\lambda_d(S)$ and $\Lambda_d(S)$ are also called "exponents of increase" of $d$ on $S$. It is implicit in Staiger's paper [37] that

$$\Lambda_{\text{comp}}(S) = 1 - \dim_{\text{comp}}(S)$$

for all $S \in \mathbf{C}$, where $\Lambda_{\text{comp}}(S)$ is defined like $\Lambda(S)$ above, but with $d$ required to be a computable martingale. Similar reasoning leads to the following characterizations of the Lyapunov exponents.

**Theorem 6.3.** *Let $S \in \mathbf{C}$ and $X \subseteq \mathbf{C}$. Then $\Lambda(S) = 1 - \dim(S)$, $\lambda(S) = 1 - \text{Dim}(S)$, $\Lambda(X) = 1 - \text{cdim}(X)$, and $\lambda(X) = 1 - \text{cDim}(X)$.*

*Proof.* We will show that $\Lambda(S) = 1 - \dim(S)$. A similar argument shows that $\lambda(S) = 1 - \text{Dim}(S)$. By Lemma 5.4, $\Lambda(X) = 1 - \text{cdim}(X)$ and $\lambda(X) = 1 - \text{cDim}(X)$ follow from the statements about sequences.

Let $t < s < \Lambda(S)$ with $t$ computable, and let $d$ be a constructive martingale for which $\lambda_d(S) > s$. Then for infinitely many $n$, $d(S[0..n-1]) > 2^{sn}$. Define a constructive $(1-t)$-gale $d'$ by $d'(w) = 2^{-t|w|}d(w)$ for all $w \in \{0,1\}^*$. Then for infinitely many $n$, we have $d'(S[0..n-1]) = 2^{-tn}d(S[0..n-1]) > 2^{(s-t)n}$, so $S \in S^\infty[d]$. Therefore $\dim(S) < 1-t$. This holds for all computable $t < \Lambda(S)$, so $\dim(S) \le 1 - \Lambda(S)$.

Let $s > \dim(S)$ be computable, and let $d$ be a constructive $s$-gale with $S \in S^\infty[d]$. Define a constructive martingale $d'$ by $d'(w) = 2^{(1-s)|w|}d(w)$ for all $w \in \{0,1\}^*$. For infinitely many $n$, we have $d(S[0..n-1]) > 1$, and for each of these $n$, $d'(S[0..n-1]) > 2^{(1-s)n}$. Therefore $\Lambda_{d'}(S) \ge 1-s$, so $\Lambda(S) \ge 1-s$. This holds for all $s > \dim(S)$, so $\Lambda(S) \ge 1 - \dim(S)$. □

Constructive strong dimension can also be used to characterize entropy rates of the type investigated by Staiger [36, 37] and Hitchcock [15].

**Definition.** *Let $A \subseteq \{0,1\}^*$.*

1. *The* entropy rate *of $A \subseteq \{0,1\}^*$ is $H_A = \limsup_{n \to \infty} \frac{\log |A_{=n}|}{n}$.*

2. *We define the sets of sequences*

$$A^{\text{i.o.}} = \{S \in \mathbf{C} | (\exists^\infty n) S[0..n-1] \in A\}$$

*and*

$$A^{\text{a.e.}} = \{S \in \mathbf{C} | (\forall^\infty n) S[0..n-1] \in A\}.$$

**Definition.** *Let $X \subseteq \mathbf{C}$. The* constructive entropy rate *of $X$ is*

$$\mathcal{H}_{\mathrm{CE}}(X) = \inf\{H_A | X \subseteq A^{\mathrm{i.o.}} \text{ and } A \in \mathrm{CE}\}$$

*and the* constructive strong entropy rate *of $X$ is*

$$\mathcal{H}_{\mathrm{CE}}^{\mathrm{str}}(X) = \inf\{H_A | X \subseteq A^{\mathrm{a.e.}} \text{ and } A \in \mathrm{CE}\}.$$

Hitchcock [15] proved that
$$\mathcal{H}_{\mathrm{CE}}(X) = \mathrm{cdim}(X) \tag{6.4}$$
for all $X \subseteq \mathbf{C}$. We have the following dual of (6.4).

**Theorem 6.4.** *For any $X \subseteq \mathbf{C}$, $\mathcal{H}_{\mathrm{CE}}^{\mathrm{str}}(X) = \mathrm{cDim}(X)$.*

*Proof.* This proof is analogous to the proof of (6.4) given in [15]. □

In the classical case, Tricot [41] has defined a set to be *regular* if its Hausdorff and packing dimensions coincide, and defined its *irregularity* to be the difference between these two fractal dimensions. Analogously, we define the c-*irregularity* (i.e., constructive irregularity) of a sequence $S \in \mathbf{C}$ to be $\mathrm{Dim}(S) - \mathrm{dim}(S)$, and we define the c-*irregularity* of a set $X \subseteq \mathbf{C}$ to be $\mathrm{cDim}(X) - \mathrm{cdim}(X)$. We define a sequence or set to be c-*regular* (i.e., *constructively regular*) if its c-irregularity is 0.

As the following result shows, the c-irregularity of a sequence may be any real number in $[0, 1]$.

**Theorem 6.5.** *For any two real numbers $0 \leq \alpha \leq \beta \leq 1$, there is a sequence $S \in \mathbf{C}$ such that $\mathrm{dim}(S) = \alpha$ and $\mathrm{Dim}(S) = \beta$.*

*Proof.* Let $R$ be a Martin-Löf random sequence. It is well-known that
$$K(R[0..n-1]) \geq n - O(1). \tag{6.5}$$

Write $R = r_1 r_2 r_3 \ldots$ where $|r_n| = 2n - 1$ for all $n$. Note that $|r_1 \cdots r_n| = n^2$.

For each $n$, define
$$\gamma_n = \begin{cases} \frac{1-\alpha}{\alpha} & \text{if } \log^* n \text{ is odd} \\ \frac{1-\beta}{\beta} & \text{if } \log^* n \text{ is even,} \end{cases}$$

and let
$$k_n = \lceil |r_n| \gamma_n \rceil.$$

We now define $S \in \mathbf{C}$ as
$$S = r_1 0^{k_1} r_2 0^{k_2} \cdots r_n 0^{k_n} \cdots .$$

Note that for all $n$,
$$\begin{aligned} |r_n 0^{k_n}| &= \lceil |r_n|(1 + \gamma_n) \rceil \\ &= \begin{cases} \lceil \frac{1}{\alpha}|r_n| \rceil & \text{if } \log^* n \text{ is odd} \\ \lceil \frac{1}{\beta}|r_n| \rceil & \text{if } \log^* n \text{ is even.} \end{cases} \end{aligned}$$

Let $w \sqsubseteq S$. Then for some $n$,
$$w = r_1 0^{k_1} \cdots r_{n-1} 0^{k_{n-1}} r_n' 0^j$$

where $r'_n \sqsubseteq r_n$ and $0 \leq j \leq k_n$. We have

$$
\begin{aligned}
K(w) &\leq K(r_1 \cdots r_{n-1} r'_n) + K(k_1) + \cdots K(k_{n-1}) + K(j) + O(1) \\
&\leq |r_1 \cdots r_{n-1} r'_n| + O(n \log n) \\
&\leq (n-1)^2 + O(n \log n).
\end{aligned}
\tag{6.6}
$$

Also,

$$
\begin{aligned}
K(r_1 \cdots r_{n-1} r'_n) &\leq K(w) + K(k_1) + \cdots + K(k_{n-1}) + K(j) + O(1) \\
&\leq K(w) + O(n \log n),
\end{aligned}
$$

so by (6.5),

$$
\begin{aligned}
K(w) &\geq K(r_1 \cdots r_{n-1} r'_n) - O(n \log n) \\
&\geq |r_1 \cdots r_{n-1} r'_n| - O(n \log n) \\
&\geq (n-1)^2 - O(n \log n).
\end{aligned}
\tag{6.7}
$$

We bound the length of $w$ in terms of $n$ as

$$
\begin{aligned}
|w| &\geq |r_1|(1 + \gamma_1) + \cdots + |r_{n-1}|(1 + \gamma_{n-1}) + |r'_n| \\
&\geq \frac{|r_1 \cdots r_{n-1}|}{\beta} \\
&= \frac{1}{\beta}(n-1)^2
\end{aligned}
\tag{6.8}
$$

and

$$
\begin{aligned}
|w| &\leq |r_1|(1 + \gamma_1) + \cdots + |r_{n-1}|(1 + \gamma_{n-1}) + |r_n|(1 + \gamma_n) + n \\
&\leq \frac{|r_1 \cdots r_{n-1} r_n|}{\alpha} + n \\
&\leq \frac{1}{\alpha}(n+1)^2.
\end{aligned}
\tag{6.9}
$$

From (6.6) and (6.8), we have

$$
\limsup_{m \to \infty} \frac{K(S[0..m-1])}{m} \leq \limsup_{n \to \infty} \frac{(n-1)^2 + O(n \log n)}{\frac{1}{\beta}(n-1)^2} = \beta,
\tag{6.10}
$$

and (6.7) and (6.9) yield

$$
\liminf_{m \to \infty} \frac{K(S[0..m-1])}{m} \geq \liminf_{n \to \infty} \frac{(n-1)^2 - O(n \log n)}{\frac{1}{\alpha}(n+1)^2} = \alpha.
\tag{6.11}
$$

For each $n$, let

$$w_n = r_1 0^{k_1} \cdots r_n 0^{k_n}.$$

Define the sequence of *towers* $t_j$ by $t_0 = 1$ and $t_{j+1} = 2^{t_j}$. If $j$ is even, then for all $t_{j-1} < i \leq t_j$,

$\gamma_i = \frac{1-\beta}{\beta}$. Then

$$|w_{t_j}| \leq t_j + \sum_{i=1}^{t_j} |r_i|(1+\gamma_i)$$
$$= t_j + \sum_{i=1}^{t_{j-1}} |r_i|(1+\gamma_i) + \frac{1}{\beta}\sum_{i=t_{j-1}+1}^{t_j} |r_i| \qquad (6.12)$$
$$\leq t_j + \frac{1}{\alpha}t_{j-1}^2 + \frac{1}{\beta}(t_j^2 - t_{j-1}^2)$$
$$\leq \frac{1}{\beta}t_j^2 + t_j + O((\log t_j)^2).$$

Similarly, if $j$ is odd, we have

$$|w_{t_j}| \geq \sum_{i=1}^{t_j} |r_i|(1+\gamma_i)$$
$$= \sum_{i=1}^{t_{j-1}} |r_i|(1+\gamma_i) + \frac{1}{\alpha}\sum_{i=t_{j-1}+1}^{t_j} |r_i| \qquad (6.13)$$
$$\geq \frac{1}{\beta}t_{j-1}{}^2 + \frac{1}{\alpha}(t_j^2 - t_{j-1}^2)$$
$$\geq \frac{1}{\alpha}t_j^2 - O((\log t_j)^2).$$

Combining (6.7) and (6.12), we have

$$\limsup_{m\to\infty} \frac{K(S[0..m-1])}{m} \geq \limsup_{n\to\infty} \frac{K(w_{t_{2n}})}{|w_{t_{2n}}|} \geq \beta. \qquad (6.14)$$

Putting (6.6) together with (6.13) yields

$$\liminf_{m\to\infty} \frac{K(S[0..m-1])}{m} \leq \liminf_{n\to\infty} \frac{K(w_{t_{2n+1}})}{|w_{t_{2n+1}}|} \leq \alpha. \qquad (6.15)$$

By (6.1), (6.11), and (6.15), we have $\dim(S) = \alpha$. By Corollary 6.2, (6.10), and (6.14), we have $\text{Dim}(S) = \beta$. □

We now come to the main theorem of this section. The following notation simplifies its statement and proof.

**Notation.** Given a bias sequence $\vec{\beta} = (\beta_0, \beta_1, \ldots)$, $n \in \mathbb{N}$, and $S \in \mathbf{C}$, let

$$H_n(\vec{\beta}) = \frac{1}{n}\sum_{i=0}^{n-1} \mathcal{H}(\beta_i),$$
$$H^-(\vec{\beta}) = \liminf_{n\to\infty} H_n(\vec{\beta}),$$
$$H^+(\vec{\beta}) = \limsup_{n\to\infty} H_n(\vec{\beta}).$$

We call $H^-(\vec{\beta})$ and $H^+(\vec{\beta})$ the *lower* and *upper average entropies*, respectively, of $\vec{\beta}$.

**Theorem 6.6.** *If $\delta \in (0, \frac{1}{2}]$ and $\vec{\beta}$ is a computable bias sequence with each $\beta_i \in [\delta, \frac{1}{2}]$, then for every sequence $R \in \text{RAND}^{\vec{\beta}}$,*

$$\dim(R) = H^-(\vec{\beta}) \text{ and } \text{Dim}(R) = H^+(\vec{\beta}).$$

Theorem 6.6 says that every sequence that is random with respect to a suitable bias sequence $\vec{\beta}$ has the lower and upper average entropies of $\vec{\beta}$ as its dimension and strong dimension, respectively. Since there exist $\vec{\beta}$-random sequences in $\Delta_2^0$ when $\vec{\beta}$ is computable, this gives a powerful and flexible method for constructing $\Delta_2^0$ sequences with given ($\Delta_2^0$-computable) dimensions and strong dimensions.

We now develop a sequence of results that are used in our proof of Theorem 6.6.

**Lemma 6.7.** *Assume that $\delta > 0$, $\epsilon > 0$, and that, for each $\beta \in [\delta, 1 - \delta]$, $\eta_\beta$ is a bounded random variable such that $\text{E}\eta_\beta \leq -\epsilon$ and $\text{E}e^{t\eta_\beta}$ is a continuous function of $\beta$ for each $t > 0$. Then there exists $\theta > 0$ such that, for all $\beta \in [\delta, 1 - \delta]$ and $t \in (0, \theta]$,*

$$\text{E}e^{t\eta_\beta} < 1 - \frac{t\epsilon}{2}.$$

*Proof.* Assume the hypothesis. Then the Dominated Convergence Theorem [3] tells us that, for all $\beta \in [\delta, 1 - \delta]$,

$$\begin{aligned}
\lim_{t \to 0^+} \frac{\text{E}e^{t\eta_\beta} - 1}{t} &= \lim_{t \to 0^+} \text{E}\frac{e^{t\eta_\beta} - 1}{t} \\
&= \text{E}\left(\lim_{t \to 0^+} \frac{e^{t\eta_\beta} - 1}{t}\right) \\
&= \text{E}\left(\eta_\beta \lim_{t \to 0^+} \frac{e^{t\eta_\beta} - 1}{t\eta_\beta}\right) \\
&= \text{E}\eta_\beta \\
&\leq -\epsilon.
\end{aligned}$$

Hence, for each $\beta \in [\delta, 1 - \delta]$, there exists $t_\beta > 0$ such that, for all $t \in (0, t_\beta]$,

$$\frac{\text{E}e^{t\eta_\beta} - 1}{t} < -\frac{3\epsilon}{4}.$$

It follows by our continuity hypothesis that, for each $\beta \in [\delta, 1 - \delta]$, there is an open neighborhood $N_\beta$ of $\beta$ such that, for all $t \in (0, t_\beta]$ and $\gamma \in N_\beta \cap [\delta, 1 - \delta]$,

$$\frac{\text{E}e^{t\eta_\gamma} - 1}{t} < -\frac{\epsilon}{2}.$$

The family $\mathcal{G} = \{N_\beta \mid \beta \in [\delta, 1-\delta]\}$ is an open cover of the compact set $[\delta, 1-\delta]$, so there is a finite set $B \subseteq [\delta, 1-\delta]$ such that the subcollection $\mathcal{G}' = \{N_\beta \mid \beta \in B\}$ is also a cover of $[\delta, 1-\delta]$. Let

$$\theta = \min\{t_\beta \mid \beta \in B\}.$$

Then $\theta > 0$ and, for all $\beta \in [\delta, 1-\delta]$ and $t \in (0, \theta]$,

$$\frac{\text{E}e^{t\eta_\beta} - 1}{t} < -\frac{\epsilon}{2},$$

whence
$$Ee^{t\eta_\beta} < 1 - \frac{t\epsilon}{2}.$$

□

**Corollary 6.8.** *For each $\delta > 0$ and $\epsilon > 0$, there exists $\theta > 0$ such that, for all $\beta \in [\delta, 1-\delta]$, if we choose $a \in \{0,1\}$ with $\mathrm{Prob}[a = 1] = \beta$, and if*
$$\eta = \xi - \mathcal{H}(\beta) - \epsilon$$
*or*
$$\eta = \mathcal{H}(\beta) - \xi - \epsilon,$$
*where*
$$\xi = (1-a)\log\frac{1}{1-\beta} + a\log\frac{1}{\beta},$$
*then*
$$Ee^{\theta\eta} < 1 - \frac{\theta\epsilon}{2}.$$

*Proof.* The random variables
$$\begin{aligned}\eta_{1,\beta} &= \xi - \mathcal{H}(\beta) - \epsilon, \\ \eta_{2,\beta} &= \mathcal{H}(\beta) - \xi - \epsilon\end{aligned}$$
satisfy the hypothesis of Lemma 6.7 with $E\eta_{1,\beta} = E\eta_{2,\beta} = -\epsilon$, so we can choose $\theta_1 > 0$ for $\eta_{1,\beta}$ and $\theta_2 > 0$ for $\eta_{2,\beta}$ as in that lemma. Letting $\theta = \min\{\theta_1, \theta_2\}$ establishes the corollary. □

**Notation.** Given a bias sequence $\vec{\beta} = (\beta_0, \beta_1, \ldots)$, $n \in \mathbb{N}$, and $S \in \mathbf{C}$, let
$$L_n(\vec{\beta})(S) = \log\frac{1}{\mu^{\vec{\beta}}(S[0..n-1])} = \sum_{i=0}^{n-1}\xi_i(S),$$
where
$$\xi_i(S) = (1-S[i])\log\frac{1}{1-\beta_i} + S[i]\log\frac{1}{\beta_i}$$
for $0 \leq i < n$.

Note that $L_n(\vec{\beta}), \xi_0, \ldots, \xi_{n-1}$ are random variables with
$$EL_n(\vec{\beta}) = \sum_{i=0}^{n-1}E\xi_i = \sum_{i=0}^{n-1}\mathcal{H}(\beta_i) = nH_n(\vec{\beta}).$$

The following large deviation theorem tells us that $L_n(\vec{\beta})$ is very unlikely to deviate significantly from this expected value.

**Theorem 6.9.** *For each $\delta > 0$ and $\epsilon > 0$, there exists $\alpha \in (0,1)$ such that, for all bias sequences $\vec{\beta} = (\beta_0, \beta_1, \ldots)$ with each $\beta_i \in [\delta, 1-\delta]$ and all $n \in \mathbb{Z}^+$, if $L_n(\vec{\beta})$ and $H_n(\vec{\beta})$ are defined as above, then*
$$P\big[|L_n(\vec{\beta}) - nH_n(\vec{\beta})| \geq \epsilon n\big] < 2\alpha^n,$$
*where the probability is computed according to $\mu^{\vec{\beta}}$.*

*Proof.* Let $\delta > 0$ and $\epsilon > 0$, and choose $\theta > 0$ as in Corollary 6.8. Let $\alpha = 1 - \frac{\theta\epsilon}{2}$, noting that $\alpha \in (0,1)$. Let $\vec{\beta}$ be as given, and let $n \in \mathbb{Z}^+$. Let $L = L_n(\vec{\beta})$, $H = H_n(\vec{\beta})$, and $\xi_0, \xi_1, \ldots$ be as above. The proof is in two parts.

1. For each $i \in \mathbb{N}$, let $\eta_i = \xi_i - \mathcal{H}(\beta_i) - \epsilon$. Then Markov's inequality, independence, and Corollary 6.8 tell us that

$$
\begin{aligned}
\mathrm{P}[L - nH \geq \epsilon n] &= P[e^{\theta(L-nH)} \geq e^{\theta\epsilon n}] \\
&\leq e^{-\theta\epsilon n} \mathrm{E} e^{\theta(L-nH)} \\
&= \mathrm{E} e^{\theta(L-nH) - \epsilon\theta n} \\
&= \mathrm{E} e^{\theta \sum_{i=0}^{n-1} \eta_i} \\
&= \mathrm{E} \prod_{i=0}^{n-1} e^{\theta\eta_i} \\
&= \prod_{i=0}^{n-1} \mathrm{E} e^{\theta\eta_i} \\
&< \alpha^n.
\end{aligned}
$$

2. Arguing as in part 1 with $\eta_i = \mathcal{H}(\beta_i) - \xi_i - \epsilon$ shows that $\mathrm{P}[nH - L \geq \epsilon n] < \alpha^n$.

By parts 1 and 2 of this proof, we now have

$$\mathrm{P}[|L - nH| \geq \epsilon n] < 2\alpha^n.$$

□

Some of our arguments are simplified by the following constructive version of a classical theorem of Kakutani [17]. Say that two bias sequences $\vec{\beta}$ and $\vec{\beta}'$ are *square-summably equivalent*, and write $\vec{\beta} \approx^2 \vec{\beta}'$, if $\sum_{i=0}^{\infty}(\beta_i - \beta_i')^2 < \infty$.

**Theorem 6.10.** (van Lambalgen [42, 43], Vovk [45]) *Let $\delta > 0$, and let $\vec{\beta}$ and $\vec{\beta}'$ be computable bias sequences with $\beta_i, \beta_i' \in [\delta, 1-\delta]$ for all $i \in \mathbb{N}$.*

1. *If $\vec{\beta} \approx^2 \vec{\beta}'$, then $\mathrm{RAND}^{\vec{\beta}} = \mathrm{RAND}^{\vec{\beta}'}$.*

2. *If $\vec{\beta} \not\approx^2 \vec{\beta}'$, then $\mathrm{RAND}^{\vec{\beta}} \cap \mathrm{RAND}^{\vec{\beta}'} = \emptyset$.*

**Corollary 6.11.** *If $\delta > 0$ and $\vec{\beta}$ is a computable bias sequence with each $\beta_i \in [\delta, 1-\delta]$, then there is an exactly computable bias sequence $\vec{\beta}'$ with each $\beta_i' \in [\frac{\delta}{2}, \beta_i]$ satisfying $\mathrm{RAND}^{\vec{\beta}'} = \mathrm{RAND}^{\vec{\beta}}$.*

*Proof.* Assume the hypothesis. Then there is a computable function $g : \mathbb{N} \times \mathbb{N} \to \mathbb{Q}$ such that $|g(i,r) - \beta_i| \leq 2^{-r}$ for all $i, r \in \mathbb{N}$. Let $m = 2 + \lceil \log \frac{1}{\delta} \rceil$, and let

$$\beta_i' = g(i, m+i) - 2^{-(m+i)}$$

for all $i \in \mathbb{N}$. It is easily verified that $\vec{\beta}'$ is exactly computable, each $\beta_i' \in [\frac{\delta}{2}, \beta_i]$, and $\vec{\beta}' \approx^2 \vec{\beta}$, whence Theorem 6.10 tells us that $\mathrm{RAND}^{\vec{\beta}'} = \mathrm{RAND}^{\vec{\beta}}$. □

**Lemma 6.12.** *If $\delta > 0$ and $\vec{\beta}$ is a computable bias sequence with each $\beta_i \in [\delta, 1-\delta]$, then every sequence $R \in \text{RAND}^{\vec{\beta}}$ satisfies*
$$L_n(\vec{\beta})(R) = nH_n(\vec{\beta}) + o(n)$$
*as $n \to \infty$.*

*Proof.* Assume the hypothesis. By Corollary 6.11, we can assume that $\vec{\beta}$ is exactly computable. Let $\epsilon > 0$. For each $n \in \mathbb{N}$, define the set
$$Y_n = \left\{ S \in \mathbf{C} \mid |L_n(\vec{\beta})(S) - nH_n(\vec{\beta})| \geq \epsilon n \right\},$$
and let
$$X_\epsilon = \{S \in \mathbf{C} \mid (\exists^\infty n) S \in Y_n\}.$$
It suffices to show that $\mu^{\vec{\beta}}_{\text{comp}}(X_\epsilon) = 0$.

For each $n \in \mathbb{N}$ and $w \in \{0,1\}^*$, let
$$d_n(w) = \begin{cases} \mu^{\vec{\beta}}(Y_n | \mathbf{C}_w) & \text{if } |w| \leq n \\ d_n(w[0..n-1]) & \text{if } |w| > n. \end{cases}$$
It is easily verified that each $d_n$ is a $\vec{\beta}$-martingale and that the function $(n,w) \mapsto d_n(w)$ is computable. It is clear that $Y_n \subseteq S^1[d_n]$ for all $n \in \mathbb{N}$. Finally, by Theorem 6.9, the series $\sum_{n=0}^\infty d_n(\lambda)$ is computably convergent, so the computable first Borel-Cantelli Lemma [22] tells us that $\mu^{\vec{\beta}}_{\text{comp}}(X_\epsilon) = 0$. □

**Lemma 6.13.** *If $\delta > 0$ and $\vec{\beta}$ is a computable bias sequence with each $\beta_i \in [\delta, \frac{1}{2}]$, then $\text{cdim}(\text{RAND}^{\vec{\beta}}) \leq H^-(\vec{\beta})$ and $\text{cDim}(\text{RAND}^{\vec{\beta}}) \leq H^+(\vec{\beta})$.*

*Proof.* Assume the hypothesis. By Corollary 6.11, we can assume that $\vec{\beta}$ is exactly computable. Let $s \in [0, \infty)$ be computable.

Define $d: \{0,1\}^* \to [0,\infty)$ by
$$d(w) = 2^{s|w|} \mu^{\vec{\beta}}(w)$$
for all $w \in \{0,1\}^*$. Then $d$ is a constructive (in fact, computable) $s$-gale. For each $R \in \mathbf{C}$ and $n \in \mathbb{N}$, if we write $z_n = R[0..n-1]$, then
$$\log d(z_n) = sn + \log \mu^{\vec{\beta}}(z_n)$$
for all $n$. In particular, if $R \in \text{RAND}^{\vec{\beta}}$, if follows by Lemma 6.12 that
$$\log d(z_n) = n[s - H_n(\vec{\beta})] + o(n) \tag{6.16}$$
as $n \to \infty$. We now verify the two parts of the lemma. For both parts, we let
$$I_\epsilon = \{n \in \mathbb{N} \mid H_n(\vec{\beta}) < s - \epsilon\}.$$

To see that $\text{cdim}(\text{RAND}^{\vec{\beta}}) \leq H^-(\vec{\beta})$, let $s > H^-(\vec{\beta})$, and let $\epsilon = \frac{s - H^-(\vec{\beta})}{2}$. Then the set $I_\epsilon$ is infinite, so (6.16) tells us that $\text{RAND}^{\vec{\beta}} \subseteq S^\infty[d]$, whence $\text{cdim}(\text{RAND}^{\vec{\beta}}) \leq s$.

To see that $\text{cDim}(\text{RAND}^{\vec{\beta}}) \leq H^+(\vec{\beta})$, let $s > H^+(\vec{\beta})$, and let $\epsilon = \frac{s - H^+(\vec{\beta})}{2}$. Then the set $I_\epsilon$ is cofinite, so (6.16) tells us that $\text{RAND}^{\vec{\beta}} \subseteq S^\infty_{\text{str}}[d]$, whence $\text{cDim}(\text{RAND}^{\vec{\beta}}) \leq s$. □

**Lemma 6.14.** *Assume that $\delta > 0$, $\vec{\beta}$ is a computable bias sequence with each $\beta_i \in [\delta, 1-\delta]$, $s \in [0, \infty)$ is computable, and $d$ is a constructive $s$-gale.*

1. *If $s < H^-(\vec{\beta})$, then $S^\infty[d] \cap \text{RAND}^{\vec{\beta}} = \emptyset$.*

2. *If $s < H^+(\vec{\beta})$, then $S^\infty_{\text{str}}[d] \cap \text{RAND}^{\vec{\beta}} = \emptyset$.*

*Proof.* Assume the hypothesis. Define $d' : \{0,1\}^* \to [0, \infty)$ by

$$d'(w) = \frac{d(w)}{2^{s|w|}\mu^{\vec{\beta}}(w)}$$

for all $w \in \{0,1\}^*$. Then $d'$ is a $\vec{\beta}$-martingale, and $d'$ is clearly constructive.

Let $R \in \text{RAND}^{\vec{\beta}}$. Then $d'$ does not succeed on $R$, so there is a constant $c > 0$ such that, for all $n \in \mathbb{N}$, if we write $z_n = R[0..n-1]$, then $d'(z_n) \leq 2^c$, whence

$$\log d(z_n) \leq c + sn + \log \mu^{\vec{\beta}}(z_n).$$

It follows by Lemma 6.12 that

$$\log d(z_n) \leq c + n[s - H_n(\vec{\beta})] + o(n)$$

as $n \to \infty$. Hence, for any $\epsilon > 0$, if we let

$$I_\epsilon = \{n \in \mathbb{Z}^+ \mid s < H_n(\vec{\beta}) - \epsilon\},$$

then $\log d(z_n) < c$ for all sufficiently large $n \in I_\epsilon$. We now verify the two parts of the lemma.

1. If $s < H^-(\vec{\beta})$, let $\epsilon = \frac{H^-(\vec{\beta})-s}{2}$. Then $I_\epsilon$ is cofinite, so $\log d(z_n) < c$ for all sufficiently large $n \in \mathbb{Z}^+$, so $R \notin S^\infty[d]$.

2. If $s < H^+(\vec{\beta})$, let $\epsilon = \frac{H^+(\vec{\beta})-s}{2}$. Then $I_\epsilon$ is infinite, so $\log d(z_n) < c$ for infinitely many $n \in \mathbb{Z}^+$, so $R \notin S^\infty_{\text{str}}[d]$.

$\square$

We now have all we need to prove the main theorem of this section.

*Proof of Theorem 6.6.* Assume the hypothesis, and let $R \in \text{RAND}^{\vec{\beta}}$. By Lemma 6.13, $\dim(R) \leq H^-(\vec{\beta})$ and $\text{Dim}(R) \leq H^+(\vec{\beta})$. To see that $\dim(R) \geq H^-(\vec{\beta})$ and $\text{Dim}(R) \geq H^+(\vec{\beta})$, let $s, t \in [0, \infty)$ be computable with $s < H^-(\vec{\beta})$ and $t < H^+(\vec{\beta})$, let $d^-$ be a constructive $s$-gale, and let $d^+$ be a constructive $t$-gale. It suffices to show that $R \notin S^\infty[d^-]$ and $R \notin S^\infty_{\text{str}}[d^+]$. But these follow immediately from Lemma 6.14 and the $\vec{\beta}$-randomness of $R$. $\square$

**Corollary 6.15.** *If $\vec{\beta}$ is a computable sequence of coin-toss biases such that $\overline{H}(\vec{\beta}) = \lim_{n \to \infty} H_n(\vec{\beta}) \in (0,1)$, then every sequence $R \in \mathbf{C}$ that is random with respect to $\vec{\beta}$ is $c$-regular, with $\dim(R) = \text{Dim}(R) = \overline{H}(\vec{\beta})$.*

Note that Corollary 6.15 strengthens Theorem 7.6 of [21] because the convergence of $H_n(\vec{\beta})$ is a weaker hypothesis than the convergence of $\vec{\beta}$.

Generalizing the construction of Chaitin's random real number $\Omega$ [6], Mayordomo [26] and, independently, Tadaki [40] defined for each $s \in (0, 1]$ and each infinite, computably enumerable set $A \subseteq \{0,1\}^*$, the real number

$$\theta_A^s = \sum \left\{ 2^{\frac{|\pi|}{s}} \,\Big|\, \pi \in \{0,1\}^* \text{ and } U(\pi) \in A \right\},$$

where $U$ is a universal self-delimiting Turing machine. Given (6.1) and Corollary 6.2 above, the following fact is implicit in Tadaki's paper.

**Theorem 6.16.** (Tadaki [40]) *For each $s \in (0,1]$ and each infinite, computably enumerable set $A \subseteq \{0,1\}^*$, the (binary expansion of the) real number $\theta_A^s$ is c-regular with $\dim(\theta_A^s) = \mathrm{Dim}(\theta_A^s) = s$.*

We define a set $X \subseteq \mathbf{C}$ to be *self-similar* if it has the form

$$X = A^\infty = \{S \in \mathbf{C} \,|\, S = w_0 w_1 w_2 \ldots \text{ for some } w_0, w_1, w_2, \ldots \in A\}$$

where is $A \subseteq \{0,1\}^*$ is a finite prefix set. Self-similar sets are examples of c-regular sets.

**Theorem 6.17.** *Let $X = A^\infty$ be self-similar where $A$ is a finite prefix set. Then $X$ is c-regular, with $\mathrm{cdim}(X) = \mathrm{cDim}(X) = \inf\{s|\sum_{w \in A} 2^{-s|w|} \leq 1\}$.*

*Proof.* We say that a string $w$ is *composite* if there are strings $w_1, \ldots, w_k \in A$ such that $w = w_1 \cdots w_k$. Let $s$ be computable such that $\sum_{w \in A} 2^{-s|w|} \leq 1$. For any computable $\epsilon > 0$ we define a constructive $(s+\epsilon)$-supergale $d$ as follows. Let $w \in \{0,1\}^*$, and let $v$ be the maximal composite proper prefix of $w$. Then

$$d(w) = \sum_{u \in A: w \sqsubseteq vu} 2^{\epsilon|w|} 2^{-s(|vu|-|w|)}.$$

For all composite strings $w$, we have $d(w) = 2^{\epsilon|w|}$. It follows that $A^\infty \subseteq S_{\mathrm{str}}^\infty[d]$, and therefore $\mathrm{cDim}(A^\infty) \leq s + \epsilon$.

Let $s$ such that $\sum_{w \in A} 2^{-s|w|} > 1$ and let $d$ be a $s$-gale. To show that $\mathrm{cdim}(A^\infty) > s$, it suffices to construct a sequence $S \in A^\infty - S^\infty[d]$. Initially, we let $w_0 = \lambda$. Assume that $w_n$ has been defined, and let $u \in A$ such that $d(w_n u) \leq d(w_n)$. We know that such a $u$ exists because of our choice of $s$. Then we let $w_{n+1} = w_n u$. Our sequence $S$ is the unique one that has $w_n \sqsubseteq S$ for all $n$. $\square$

Dai, Lathrop, Lutz, and Mayordomo [7] investigated the *finite-state compression ration* $\rho_{\mathrm{FS}}(S)$, defined for each sequence $S \in \mathbf{C}$ to be the infimum, taken over all information-lossless finite-state compressors $C$ (a model defined in Shannon's 1948 paper [35]) of the *(lower) compression ratio*

$$\rho_C(S) = \liminf_{n \to \infty} \frac{|C(S[0..n-1])|}{n}.$$

They proved that

$$\rho_{\mathrm{FS}}(S) = \dim_{\mathrm{FS}}(S) \tag{6.17}$$

for all $S \in \mathbf{C}$. However, it has been pointed out that the compression ratio $\rho_{\mathrm{FS}}(S)$ differs from the one investigated by Ziv [46]. Ziv was instead concerned with the ratio $R_{\mathrm{FS}}(S)$ defined by

$$R_{\mathrm{FS}}(S) = \inf_{k \in \mathbb{N}} \limsup_{n \to \infty} \inf_{C \in \mathcal{C}_k} \frac{|C(S[0..n-1])|}{n},$$

where $\mathcal{C}_k$ is the set of all $k$-state information-lossless finite-state compressors. The following result, together with (6.17), clarifies the relationship between $\rho_{\text{FS}}(S)$ and $R_{\text{FS}}(S)$.

**Theorem 6.18.** *For all $S \in \mathbf{C}$, $R_{\text{FS}}(S) = \text{Dim}_{\text{FS}}(S)$.*

The proof of Theorem 6.18 is based on the following lemma.

**Lemma 6.19.** *Let $\mathcal{C}$ be the set of all finite-state compressors. For all $S \in \mathbf{C}$,*

$$R_{\text{FS}}(S) = \inf_{C \in \mathcal{C}} \limsup_{n \to \infty} \frac{|C(S[0..n-1])|}{n}.$$

*Proof.* Let

$$R'_{\text{FS}}(S) = \inf_{C \in \mathcal{C}} \limsup_{n \to \infty} \frac{|C(S[0..n-1])|}{n}.$$

The inequality $R_{\text{FS}}(S) \leq R'_{\text{FS}}(S)$ is trivial. We use several results from [7] to obtain for each $k \in \mathbb{N}$ and $\epsilon > 0$ a finite-state compressor $C_{k,\epsilon}$ that is nearly optimal for all compressors in $\mathcal{C}_k$. From Lemma 7.7 in [7] we obtain a finite-state gambler for each $C \in \mathcal{C}_k$. By Lemma 3.7 in [7], we can combine these gamblers into a single finite-state gambler. Theorem 4.5 and Lemma 3.11 in [7] convert this single gambler into a 1-account nonvanishing finite-state gambler and finally Lemma 7.10 converts this to the finite-state compressor $C_{k,\epsilon}$. Combining the five cited constructions in [7] we obtain that there is a constant $c_{k,\epsilon}$ such that for all $w \in \{0,1\}^*$ and $C \in \mathcal{C}_k$,

$$|C_{k,\epsilon}(w)| \leq |C(w)| + \epsilon|w| + c_{k,\epsilon}.$$

Then for all $k \in \mathbb{N}$ and $\epsilon > 0$,

$$\begin{aligned} R'_{\text{FS}}(S) &\leq \limsup_{n \to \infty} \frac{|C_{k,\epsilon}(S[0..n-1])|}{n} \\ &\leq \limsup_{n \to \infty} \inf_{C \in \mathcal{C}_k} \frac{|C(S[0..n-1])|}{n} + \epsilon, \end{aligned}$$

so $R'_{\text{FS}}(S) \leq R_{\text{FS}}(S)$. $\square$

*Proof of Theorem 6.18.* The equality

$$\text{Dim}_{\text{FS}}(S) = \inf_{C \in \mathcal{C}} \limsup_{n \to \infty} \frac{|C(S[0..n-1])|}{n}$$

has a proof analogous to that of (6.17) given in [7]. Together with Lemma 6.19, this implies that $R_{\text{FS}}(S) = \text{Dim}_{\text{FS}}(S)$. $\square$

Thus, mathematically, the compression ratios $\rho_{\text{FS}}(S)$ and $R_{\text{FS}}(S)$ are both natural: they are the finite-state effectivizations of the Hausdorff and packing dimensions, respectively.

# 7 Computational Complexity

In this section we prove our third main theorem, which says that the dimensions and strong dimensions of polynomial-time many-one degrees in exponential time are essentially unrestricted. Our proof of this result uses convenient characterizations of p-dimension and strong p-dimension in terms of feasible unpredictability.

**Definition.** *A* predictor *is a function* $\pi : \{0,1\}^* \times \{0,1\} \to [0,1]$ *such that for all* $w \in \{0,1\}^*$, $\pi(w,0) + \pi(w,1) = 1$.

We interpret $\pi(w,b)$ as the predictor's estimate of the probability that the bit $b$ will occur next given that $w$ has occurred. We write $\Pi(p)$ for the class of all feasible predictors.

**Definition.** *Let* $w \in \{0,1\}^*$, $S \in \mathbf{C}$, *and* $X \subseteq \mathbf{C}$.

1. *The* cumulative log-loss *of $\pi$ on $w$ is*

$$\mathcal{L}^{\log}(\pi, w) = \sum_{i=0}^{|w|-1} \log \frac{1}{\pi(w[0..i-1], w[i])}.$$

2. *The* log-loss rate *of $\pi$ on $S$ is*

$$\mathcal{L}^{\log}(\pi, S) = \liminf_{n \to \infty} \frac{\mathcal{L}^{\log}(\pi, S[0..n-1])}{n}.$$

3. *The* strong log-loss rate *of $\pi$ on $S$ is*

$$\mathcal{L}^{\log}_{\mathrm{str}}(\pi, S) = \limsup_{n \to \infty} \frac{\mathcal{L}^{\log}(\pi, S[0..n-1])}{n}.$$

4. *The* (worst-case) log-loss *of $\pi$ on $X$ is*

$$\mathcal{L}^{\log}(\pi, X) = \sup_{S \in X} \mathcal{L}^{\log}(\pi, S).$$

5. *The* (worst-case) strong log-loss *of $\pi$ on $X$ is*

$$\mathcal{L}^{\log}_{\mathrm{str}}(\pi, X) = \sup_{S \in X} \mathcal{L}^{\log}_{\mathrm{str}}(\pi, S).$$

6. *The* feasible log-loss unpredictability *of $X$ is*

$$\mathrm{unpred}^{\log}_{\mathrm{p}}(X) = \inf_{\pi \in \Pi(\mathrm{p})} \mathcal{L}^{\log}(\pi, X).$$

7. *The* feasible strong log-loss unpredictability *of $X$ is*

$$\mathrm{Unpred}^{\log}_{\mathrm{p}}(X) = \inf_{\pi \in \Pi(\mathrm{p})} \mathcal{L}^{\log}_{\mathrm{str}}(\pi, X).$$

Hitchcock [13] showed that feasible dimension exactly characterizes feasible log-loss unpredictability, that is,
$$\text{unpred}_{\text{p}}^{\log}(X) = \dim_{\text{p}}(X) \tag{7.1}$$
for all $X \subseteq \mathbf{C}$. We have the following dual result for strong dimension.

**Theorem 7.1.** *For all $X \subseteq \mathbf{C}$, $\text{Unpred}_{\text{p}}^{\log}(X) = \text{Dim}_{\text{p}}(X)$.*

The following theorem is the main result of this section. This theorem and its proof are motivated by analogous, but simpler arguments by Ambos-Spies, Merkle, Reimann and Stephan [1].

**Theorem 7.2.** *For every pair of $\Delta_2^0$-computable real numbers $x, y$ with $0 \leq x \leq y \leq 1$, there exists $A \in \text{E}$ such that*
$$\dim_{\text{p}}(\deg_{\text{m}}^{\text{P}}(A)) = \dim(\deg_{\text{m}}^{\text{P}}(A)|\text{E}) = x$$
*and*
$$\text{Dim}_{\text{p}}(\deg_{\text{m}}^{\text{P}}(A)) = \text{Dim}(\deg_{\text{m}}^{\text{P}}(A)|\text{E}) = y.$$

We now develop the proof of Theorem 7.2. Our proof uses several preliminary results.

The first part of the following theorem is due to Ambos-Spies, Merkle, Reimann and Stephan [1]. The second part is an exact dual of the first part.

**Theorem 7.3.** *Let $A \in \text{E}$.*

1. *$\dim_{\text{p}}(\deg_{\text{m}}^{\text{P}}(A)) = \dim_{\text{p}}(\text{P}_{\text{m}}(A))$ and $\dim(\deg_{\text{m}}^{\text{P}}(A)|\text{E}) = \dim(\text{P}_{\text{m}}(A)|\text{E})$.*

2. *$\text{Dim}_{\text{p}}(\deg_{\text{m}}^{\text{P}}(A)) = \text{Dim}_{\text{p}}(\text{P}_{\text{m}}(A))$ and $\text{Dim}(\deg_{\text{m}}^{\text{P}}(A)|\text{E}) = \text{Dim}(\text{P}_{\text{m}}(A)|\text{E})$.*

The following lemma is a time-bounded version of Lemma 6.14.

**Lemma 7.4.** *Assume that $k, l \in \mathbb{Z}^+$, $\delta > 0$, $\vec{\beta}$ is an exactly $n^l$-time-computable bias sequence with each $\beta_i \in \mathbb{Q} \cap [\delta, 1-\delta]$, $s \in \mathbb{Q} \cap [0, \infty)$, and $d$ is an $n^k$-time-computable $s$-gale.*

1. *If $s < H^-(\vec{\beta})$, then $S^\infty[d] \bigcap \text{RAND}^{\vec{\beta}}(n^{k+2l+1}) = \emptyset$.*

2. *If $s < H^+(\vec{\beta})$, then $S^\infty_{\text{str}}[d] \bigcap \text{RAND}^{\vec{\beta}}(n^{k+2l+1}) = \emptyset$.*

*Proof.* We proceed exactly as in the proof of Lemma 6.14, noting that our present hypothesis implies that the $\vec{\beta}$-martingale $d'$ is $O(n^{k+2l+1})$-time-computable. □

Our proof of Theorem 7.2 also uses the *martingale dilation technique*, which was introduced by Ambos-Spies, Terwijn, and Zheng [2] and extended by Breutzmann and Lutz [4].

**Definition.** *The restriction of a string $w \in \{0,1\}^*$ to a language $A \subseteq \{0,1\}^*$ is the string $w \restriction A$ defined by the following recursion.*

1. $\lambda \restriction A = \lambda$.

2. *For $w \in \{0,1\}^*$ and $b \in \{0,1\}$,*
$$(wb) \restriction A = \begin{cases} (w \restriction A)b & \text{if } s_{|w|} \in A, \\ w \restriction A & \text{if } s_{|w|} \notin A. \end{cases}$$

(That is, $w \upharpoonright A$ is the concatenation of the successive bits $w[i]$ for which $s_i \in A$.)

**Definition.** *A function $f : \{0,1\}^* \longrightarrow \{0,1\}^*$ is strictly increasing if, for all $x, y \in \{0,1\}^*$,*

$$x < y \Longrightarrow f(x) < f(y),$$

*where $<$ is the standard ordering of $\{0,1\}^*$.*

**Notation.** If $f : \{0,1\}^* \longrightarrow \{0,1\}^*$, then for each $n \in \mathbb{N}$, let $n_f$ be the unique integer such that $f(s_n) = s_{n_f}$.

**Definition.** *If $f : \{0,1\}^* \longrightarrow \{0,1\}^*$ is strictly increasing and $\vec{\beta}$ is a bias sequence, then the $f$-dilation of $\vec{\beta}$ is the bias sequence $\vec{\beta}^f$ given by $\beta_n^f = \beta_{n_f}$ for all $n \in \mathbb{N}$.*

**Observation 7.5.** *If $f : \{0,1\}^* \longrightarrow \{0,1\}^*$ is strictly increasing and $A \subseteq \{0,1\}^*$, then for all $n \in \mathbb{N}$,*

$$\chi_{f^{-1}(A)}[0..n-1] = \chi_A[0..n_f - 1] \upharpoonright range(f).$$

**Definition.** *If $f : \{0,1\}^* \longrightarrow \{0,1\}^*$ is strictly increasing and $d$ is a martingale, then then the $f$-dilation of $d$ is the function $f\hat{\ }d : \{0,1\}^* \longrightarrow [0, \infty)$,*

$$f\hat{\ }d(w) = d(w \upharpoonright range(f)).$$

Intuitively, the $f$-dilation of $d$ is a strategy for betting on a language $A$, assuming that $d$ itself is a good betting strategy for betting on the language $f^{-1}(A)$. Given an opportunity to bet on the membership of a string $y = f(x)$ in A, $f\hat{\ }d$ bets exactly as $d$ would bet on the membership or nonmembership of $x$ in $f^{-1}(A)$.

The following result is a special case of Theorem 6.3 in [4].

**Theorem 7.6.** *(Martingale Dilation Theorem - Breutzmann and Lutz [4]) Assume that $\vec{\beta}$ is a bias sequence with each $\beta_i \in (0,1)$, $f : \{0,1\}^* \longrightarrow \{0,1\}^*$ is strictly increasing, and $d$ is a $\vec{\beta}^f$-martingale. Then $f\hat{\ }d$ is a $\vec{\beta}$-martingale and, for every language $A \subseteq \{0,1\}^*$, if $d$ succeeds on $f^{-1}(A)$, then $f\hat{\ }d$ succeeds on A.*

**Notation.** For each $k \in \mathbb{Z}^+$, define $g_k : \{0,1\}^* \longrightarrow \{0,1\}^*$ by $g_k(x) = 0^{|x|^k}1x$. Note that each $g_k$ is strictly increasing and computable in polynomial time.

**Lemma 7.7.** *Assume that $\vec{\beta}$ is a bias sequence with each $\beta_i \in (0,1)$, and $R \in \mathrm{RAND}^{\vec{\beta}}(n^2)$. Then, for each $k \geq 2$, $g_k^{-1}(R) \in \mathrm{RAND}^{\vec{\alpha}}(n^k)$, where $\vec{\alpha} = \vec{\beta}^{g_k}$.*

*Proof.* Let $\vec{\beta}$, $k$, and $\vec{\alpha}$ be as given, and assume that $g_k^{-1}(R) \notin \mathrm{RAND}^{\vec{\alpha}}(n^k)$. Then there is an $n^k$-time-computable $\vec{\alpha}$-martingale $d$ that succeeds on $g_k^{-1}(R)$. It follows by Theorem 7.6 that $g_k\hat{\ }d$ is a $\vec{\beta}$-martingale that succeeds on $R$. The time required to compute $g_k\hat{\ }d(w)$ is $\mathrm{O}(|w|^2 + |w'|^k)$ steps, where $w' = w \upharpoonright range(g_k)$. (This allows $\mathrm{O}(|w|^2)$ steps to compute $w'$ and then $\mathrm{O}(|w|^k)$ steps to compute $d(w')$.) Now $|w'|$ is bounded above by the number of strings $x$ such that $|x|^k + |x| + 1 \leq |s_{|w|}| = \lfloor \log(1 + |w|) \rfloor$, so $|w'| \leq 2^{1+\log(1+|w|)^{\frac{1}{k}}}$, so the time required to compute $g_k\hat{\ }d(w)$ is

$$\mathrm{O}(|w|^2 + 2^k 2^{k(\log(1+|w|))^{\frac{1}{k}}}) = \mathrm{O}(|w|^2)$$

steps. Thus $g_k\hat{\ }d(w)$ is is an $n^2$-time computable $\vec{\beta}$-martingale, so $R \notin \mathrm{RAND}^{\vec{\beta}}(n^2)$. □

**Notation.** From here through the proof of Theorem 7.2, we assume that $\alpha$ and $\beta$ are $\Delta_2^0$-computable real numbers with $0 \leq \alpha \leq \beta \leq 1/2$. It is well-known that a real number is $\Delta_2^0$-computable if and only if there is a computable sequence of rationals that converge to it. Slowing down this construction gives polynomial-time functions $\hat{\alpha}, \hat{\beta} : \mathbb{N} \to \mathbb{Q}$ such that $\lim_{n \to \infty} \hat{\alpha}(n) = \alpha$ and $\lim_{n \to \infty} \hat{\beta}(n) = \beta$. We also assume that $\frac{1}{n} \leq \hat{\alpha}(n) \leq \hat{\beta}(n)$ for all $n$. For each $n$, we let

$$\kappa(n) = \begin{cases} \hat{\alpha}(n) & \text{if } n \text{ is even} \\ \hat{\beta}(n) & \text{if } n \text{ is odd} \end{cases}$$

and define a special-purpose bias sequence $\vec{\gamma}$ by

$$\gamma_n = \kappa(\log^* n).$$

Note that $\vec{\gamma}$ is $O(n)$-time-computable, $\frac{1}{\log^* n} \leq \gamma_n$ for all $n$, $H^-(\vec{\gamma}) = \mathcal{H}(\alpha)$, and $H^+(\vec{\gamma}) = \mathcal{H}(\beta)$.

We now use the unpredictability characterizations from the beginning of this section to establish upper bounds on the dimensions and strong dimensions of lower spans of sequences random relative to $\vec{\gamma}$.

**Lemma 7.8.** *For each* $R \in \mathrm{RAND}^{\vec{\gamma}}(n^5)$,

$$\dim_\mathrm{p}(\mathrm{P_m}(R)) \leq \mathcal{H}(\alpha)$$

*and*

$$\mathrm{Dim}_\mathrm{p}(\mathrm{P_m}(R)) \leq \mathcal{H}(\beta).$$

*Proof.* For now, fix a polynomial-time function $f : \{0,1\}^* \to \{0,1\}^*$. The *collision set* of $f$ is

$$C_f = \{j \mid (\exists i < j) f(s_i) = f(s_j)\}.$$

For each $n \in \mathbb{N}$, let

$$\#C_f(n) = |C_f \cap \{0, \ldots, n-1\}|.$$

We use $f$ to define the predictors

$$\pi_0^f(w, b) = \begin{cases} \frac{1}{2} & \text{if } |w| \notin C_f \\ w[i]^b(1 - w[i])^{1-b} & \text{if } |w| \in C_f \text{ and } i = \min\{j \mid f(s_j) = f(s_{|w|})\} \end{cases}$$

and

$$\pi_1^f(w, b) = \begin{cases} (\gamma_n^f)^b(1 - \gamma_n^f)^{1-b} & \text{if } |w| \notin C_f \\ w[i]^b(1 - w[i])^{1-b} & \text{if } |w| \in C_f \text{ and } i = \min\{j \mid f(s_j) = f(s_{|w|})\} \end{cases}$$

for all $w \in \{0,1\}^*$ and $b \in \{0,1\}$.

For each $S \in \mathbf{C}$, we now define several objects to facilitate the proof. First, we let

$$A^f(S) = f^{-1}(S);$$

that is, $A^f(S)$ is the language $\leq_\mathrm{m}^\mathrm{P}$-reduced to $S$ by $f$. Observe that for all $w \sqsubseteq A^f(S)$,

$$\mathcal{L}^{\log}(\pi_0^f, w) = |w| - \#C_f(|w|). \tag{7.2}$$

Define the sequence of *towers* $t_j$ by $t_0 = 1$ and $t_{j+1} = 2^{t_j}$. For any $j \in \mathbb{N}$ and $t_j < n \leq t_{j+1}$, define the entropy quantity

$$H_n^f = \sum_{\substack{i < n \\ i \notin C_f \text{ and } i_f > t_{j-1}}} \mathcal{H}(\gamma_n^f)$$

and the random variable

$$L_n^f(S) = \sum_{\substack{i < n \\ i \notin C_f \text{ and } i_f > t_{j-1}}} \log \frac{1}{\pi_1^f(A^f(S)[0..i-1]), A^f(S)[i])}.$$

(Recall that $i_f$ is the unique number such that $f(s_i) = s_{i_f}$.) We have

$$\begin{aligned}
\mathcal{L}^{\log}(\pi_1^f, A^f(S)[0..n-1]) &= \sum_{i<n} \log \frac{1}{\pi_1^f(A^f(S[0..i-1]), A^f(S[i]))} \\
&= \sum_{\substack{i<n \\ i \notin C_f}} \log \frac{1}{\pi_1^f(A^f(S[0..i-1]), A^f(S[i]))} \\
&= L_n^f(S) + \sum_{\substack{i<n \\ i \notin C_f \text{ and } i_f \leq t_{j-1}}} \log \frac{1}{\pi_1^f(A^f(S[0..i-1]), A^f(S[i]))} \quad (7.3) \\
&\leq L_n^f(S) + \sum_{\substack{i<n \\ i \notin C_f \text{ and } i_f \leq t_{j-1}}} \log \log^* i_f \\
&\leq L_n^f(S) + (t_{j-1} + 1) \log(j-1) \\
&\leq L_n^f(S) + (1 + \log n) \log^* n,
\end{aligned}$$

for all $n$. (Here we used the fact that $\gamma_i \geq \frac{1}{\log^* i}$ for all $i$.) Finally, for any $\epsilon > 0$ and $\theta \in (0,1)$, define the set

$$J_{\theta,\epsilon}^f(S) = \{n \mid \#C_f(n) < (1-\theta)n \text{ and } L_n^f(S) \geq H_n^f + \epsilon n\}$$

of natural numbers.

**Claim.** *For any rational $\theta \in (0,1)$ and $\epsilon > 0$,*

$$\mu_{n^5}^{\vec{\gamma}}\left(\{S \mid J_{\theta,\epsilon}^f(S) \text{ is finite}\}\right) = 1.$$

*Proof of Claim.* The argument is similar to the proof of Lemma 6.12. For each $n \in \mathbb{N}$, define the set

$$Y_n = \begin{cases} \emptyset & \text{if } \#C_f(n) \geq (1-\theta)n \\ \{S \mid L_n^f(S) \geq H_n^f + \epsilon n\} & \text{otherwise,} \end{cases}$$

and let

$$X_\epsilon = \{S \in \mathbf{C} \mid (\exists^\infty n) S \in Y_n\}.$$

To prove the claim, we will show that $\mu_{n^5}^{\vec{\gamma}}(X_\epsilon) = 0$.

For each $n \in \mathbb{N}$ and $w \in \{0,1\}^*$, let

$$d_n(w) = \begin{cases} \mu^{\vec{\gamma}}(Y_n | \mathbf{C}_w) & \text{if } |w| \leq n \\ d_n(w[0..n-1]) & \text{if } |w| > n. \end{cases}$$

It is clear that each $d_n$ is a $\vec{\gamma}$-martingale and that $Y_n \subseteq S^1[d_n]$ for all $n \in \mathbb{N}$.

Let $S \in \mathbf{C}$. For each $n, j \in \mathbb{N}$, let

$$I_j^n = \{i_f \mid i < n, i \notin C_f, \text{ and } \log^* i_f = j\}.$$

Also, define $S^+ = \{i \mid S[i] = 1\}$ and $S^- = \{i \mid S[i] = 0\}$. Then, if $n$ is large enough to ensure that $\log^* i_f \leq 1 + \log^* n$ for all $i < n$, we have

$$L_n^f(S) = \sum_{k=(\log^* n)-1}^{(\log^* n)+1} |I_k^n \cap S^+| \log \frac{1}{\kappa(k)} + |I_k^n \cap S^-| \log \frac{1}{1-\kappa(k)}.$$

For any $n$ and $k$, write $i(n,k) = |I_k^n|$. Let $T_n$ be the set of all tuples $(l_{-1}, l_0, l_1)$ satisfying $0 \leq l_r \leq i(n, j+r)$ for $-1 \leq r \leq 1$ and

$$\sum_{r=-1}^{1} l_r \log \frac{1}{\kappa(j+r)} + (i(n,j+r) - l_r) \log \frac{1}{1-\kappa(j+r)} \geq H_n^f + \epsilon n,$$

where $j = \log^* n$. Then we have

$$\mu^{\vec{\gamma}}(Y_n) = \sum_{(l_{-1}, l_0, l_1) \in T_n} \prod_{r=-1}^{1} \binom{i(n, j+r)}{l_r} \kappa(j+r)^{l_r} (1 - \kappa(j+r))^{i(n,j+r)-l_r}.$$

We can write a similar formula for $\mu^{\vec{\gamma}}(Y_n | \mathbf{C}_w)$ when $w \neq \lambda$. From this it follows that the mapping $(n, w) \mapsto d_n(w)$ is exactly computable in $O(n^3)$ time.

By Theorem 6.9, there exists $\delta \in (0, 1)$ such that for all $n \in \mathbb{N}$ with $Y_n \neq \emptyset$, we have

$$\mu^{\vec{\gamma}}(Y_n) < 2\delta^{n - \#C_f(n)} < 2\delta^{\theta n}.$$

It follows that the series $\sum_{n=0}^{\infty} d_n(\lambda)$ is p-convergent, so a routine extension of the polynomial-time first Borel-Cantelli Lemma [22] tells us that $\mu_{n^5}^{\vec{\gamma}}(X_\epsilon) = 0$. □ Claim.

Let $R \in \text{RAND}^{\vec{\gamma}}(n^5)$. Let $\epsilon > 0$ and $\theta < \mathcal{H}(\alpha)$ be rational. Then by the above claim, $J_{\theta,\epsilon}^f(R)$ is finite. That is, for all but finitely many $n$,

$$\#C_f(n) \geq (1-\theta)n \text{ or } L_n^f(R) < H_n^f + \epsilon n. \tag{7.4}$$

Writing $w_n = A^f(R)[0..n-1]$, (7.4) combined with (7.2) and (7.3) implies that

$$\mathcal{L}^{\log}(\pi_0^f, w_n) \leq \theta n < \mathcal{H}(\alpha) n \tag{7.5}$$

or

$$\mathcal{L}^{\log}(\pi_1^f, w_n) < H_n^f + \epsilon n + (1 + \log n) \log^* n. \tag{7.6}$$

As
$$\limsup_{n\to\infty} \frac{H_n^f}{n} \leq \mathcal{H}(\beta),$$
it follows that
$$\limsup_{n\to\infty} \frac{\min\{\mathcal{L}^{\log}(\pi_0^f, w_n), \mathcal{L}^{\log}(\pi_1^f, w_n)\}}{n} \leq \mathcal{H}(\beta) + \epsilon. \tag{7.7}$$
If (7.5) holds for infinitely many $n$, then
$$\mathcal{L}^{\log}(\pi_0^f, A^f(R)) \leq \mathcal{H}(\alpha). \tag{7.8}$$
Otherwise, (7.6) holds for almost all $n$. Assuming
$$\liminf_{n\to\infty} \frac{H_n^f}{n} \leq \mathcal{H}(\alpha), \tag{7.9}$$
in this case we have
$$\mathcal{L}^{\log}(\pi_1^f, A^f(R)) \leq \mathcal{H}(\alpha) + \epsilon. \tag{7.10}$$
We now verify (7.9). For each $n$, let $m(n) = t_n^2$. Then for sufficiently large $n$, we have $i_f < t_{n+1}$ for all $i < m(n)$. Using the sets $I_n^k$ from the proof of the claim, we then have
$$\begin{aligned} H_{m(n)}^f &= \left|I_n^{m(n)}\right|\mathcal{H}(\kappa(n)) + \left|I_{n+1}^{m(n)}\right|\mathcal{H}(\kappa(n+1)) \\ &\leq (t_n+1)\mathcal{H}(\kappa(n)) + m(n)\mathcal{H}(\kappa(n+1)). \end{aligned}$$
As $t_n = o(m(n))$ and $\kappa(2n) \to \alpha$ as $n \to \infty$, we have
$$\liminf_{n\to\infty} \frac{H_n^f}{n} \leq \liminf_{n\to\infty} \frac{H_{m(2n+1)}^f}{m(2n+1)} \leq \mathcal{H}(\alpha).$$

For each polynomial-time reduction $f$, we have defined and analyzed two predictors $\pi_0^f$ and $\pi_1^f$. We now show how to combine all these predictors into a single predictor that will establish the lemma.

Let $\{f_j \mid j \in \mathbb{N}\}$ be a uniform enumeration of all polynomial-time functions $f_j : \{0,1\}^* \to \{0,1\}^*$ such that $f_j(x)$ is computable in $O(2^{|x|} + j)$ steps. For any predictor $\rho$, define a probability measure $\mu[\rho]$ by
$$\mu[\rho](w) = \prod_{i=0}^{|w|-1} \rho(w[0..i-1], w[i])$$
for all $w \in \{0,1\}^*$. For each $m \in \mathbb{N}$ and $w \in \{0,1\}^m$, let
$$\mu_m(w) = 2^{-(2m+1)} + \sum_{j=0}^{m-1} 2^{-(2j+3)}\left(\mu[\pi_0^{f_j}](w) + \frac{1}{2}\mu[\pi_1^{f_j}](w)\right).$$

Then

$$\begin{aligned}
\mu_{m+1}(w0) + \mu_{m+1}(w1) &= 2^{-(2m+3)} + \sum_{j=0}^{m} 2^{-(2j+3)} \left( \mu[\pi_0^{f_j}](w0) + \frac{1}{2}\mu[\pi_1^{f_j}](w0) \right) \\
&\quad + 2^{-(2m+3)} + \sum_{j=0}^{m} 2^{-(2j+3)} \left( \mu[\pi_0^{f_j}](w1) + \frac{1}{2}\mu[\pi_1^{f_j}](w1) \right) \\
&= 2^{-(2m+2)} + \sum_{j=0}^{m} 2^{-(2j+3)} \left( \mu[\pi_0^{f_j}](w) + \frac{1}{2}\mu[\pi_1^{f_j}](w) \right) \\
&= 2^{-(2m+3)} \left( 2 + \mu[\pi_0^{f_m}](w) + \frac{1}{2}\mu[\pi_1^{f_m}](w) \right) + \mu_m(w) - 2^{-(2m+1)} \\
&\leq \mu_m(w) + 2^{-(2m+3)} \left( 3 + \frac{1}{2} \right) - 2^{-(2m+1)} \\
&< \mu_m(w).
\end{aligned}$$

Now define a predictor $\pi$ by

$$\begin{aligned}
\pi(w, 1) &= \frac{\mu_{|w|+1}(w1)}{\mu_{|w|}(w)} \\
\pi(w, 0) &= 1 - \pi(w, 1).
\end{aligned}$$

Then for all $w \in \{0,1\}^*$ and $b \in \{0,1\}$,

$$\pi(w, b) \geq \frac{\mu_{|w|+1}(wb)}{\mu_{|w|}(w)}.$$

For all $w \in \{0,1\}^*$, $i \in \{0,1\}$, and $j < |w|$, we have

$$\begin{aligned}
\mathcal{L}^{\log}(\pi, w) &= \sum_{i=0}^{|w|-1} \log \frac{1}{\pi(w[0..i-1], w[i])} \\
&\leq \sum_{i=0}^{|w|-1} \log \frac{\mu_i(w[0..i-1])}{\mu_{i+1}(w[0..i])} \\
&= \log \frac{\mu_0(\lambda)}{\mu_{|w|}(w)} \\
&\leq \log \frac{2^{2j+3+i}}{\mu[\pi_i^{f_j}](w)} \\
&= 2j + 3 + i + \mathcal{L}^{\log}(\pi_i^{f_j}, w).
\end{aligned}$$

For any $j \in \mathbb{N}$, it follows that

$$\mathcal{L}^{\log}_{\text{str}}(\pi, A^{f_j}(R)) \leq \mathcal{H}(\beta) + \epsilon$$

by using $f = f_j$ in (7.7). Also, since either (7.8) or (7.10) holds for $f = f_j$, we have

$$\mathcal{L}^{\log}(\pi, A^{f_j}(R)) \leq \mathcal{H}(\alpha) + \epsilon.$$

As $\pi$ is (exactly) polynomial-time computable, this establishes that

$$P_m(R) = \{A^{f_j}(R) \mid j \in \mathbb{N}\}$$

has p-dimension at most $\mathcal{H}(\alpha) + \epsilon$ by (7.1) and strong p-dimension at most $\mathcal{H}(\beta) + \epsilon$ by Theorem 7.1. As $\epsilon > 0$ was arbitrary, the lemma follows. □

*Proof of Theorem 7.2.* in the case $x = \mathcal{H}(\alpha)$, $y = \mathcal{H}(\beta)$, where $\alpha, \beta \in \mathbb{Q}$ and $0 < \alpha \leq \beta \leq 1/2$.

By a routine diagonalization, there is a language $A \in \text{RAND}^{\vec{\gamma}}(n^5) \cap \text{E}$. By Theorem 7.3, it suffices to prove that

$$\dim_p(P_m(A)) = \dim(P_m(A)|E) = \mathcal{H}(\alpha)$$

and

$$\text{Dim}_p(P_m(A)) = \text{Dim}(P_m(A)|E) = \mathcal{H}(\beta).$$

By Lemma 7.8, then, it suffices to prove that

$$\dim(P_m(A)|E) \geq \mathcal{H}(\alpha)$$

and

$$\text{Dim}(P_m(A)|E) \geq \mathcal{H}(\beta).$$

To see that these hold, let $s, t \in \mathbb{Q}$ with $s < \mathcal{H}(\alpha)$ and $t < \mathcal{H}(\beta)$, let $k \in \mathbb{Z}^+$, let $d^-$ be an $n^k$-time computable $s$-gale, and let $d^+$ be an $n^k$-time computable $t$-gale. It suffices to show that

$$P_m(A) \cap E \not\subseteq S^\infty[d^-] \tag{7.11}$$

and

$$P_m(A) \cap E \not\subseteq S^\infty_{\text{str}}[d^+] \tag{7.12}$$

Let $B = g_{k+3}^{-1}(A)$. It is clear that $B \in P_m(A) \cap E$. Also, by Lemma 7.7, $B \in \text{RAND}^{\vec{\gamma}'}(n^k)$, where $\vec{\gamma}' = \vec{\gamma}^{g_{k+3}}$. Since

$$s < \mathcal{H}(\alpha) = H^-(\vec{\gamma}) = H^-(\vec{\gamma}')$$

and

$$t < \mathcal{H}(\beta) = H^+(\vec{\gamma}) = H^+(\vec{\gamma}')$$

and $\vec{\gamma}'$ is $O(n)$-time-computable, Lemma 7.4 tells us that $B \notin S^\infty[d^-]$ and $B \notin S^\infty_{\text{str}}[d^+]$. Thus 7.11 and 7.12 hold.

□

In light of Theorem 7.2, the following question concerning the relativized feasible dimension of NP is natural.

**Open Question 7.9.** *For which pairs of real numbers $\alpha, \beta \in [0,1]$ does there exist an oracle $A$ such that $\dim_{p^A}(\text{NP}^A) = \alpha$ and $\text{Dim}_{p^A}(\text{NP}^A) = \beta$?*

We conclude this section with two brief observations.

Fortnow and Lutz [11] have recently established a tight quantitative relationship between p-dimension and feasible predictability. Specifically, for each $X \subseteq \mathbf{C}$, they investigated the quantity $\text{Pred}_p(X)$ which is the supremum, for all feasible predictors $\pi$, of the *(worst-case, upper) success rate*

$$\pi^+(S) = \inf_{S \in X} \limsup_{n \to \infty} \pi^+(S[0..n-1]) \tag{7.13}$$

where
$$\pi^+(w) = \sum_{i=0}^{|w|-1} \pi(w[0..i-1], w[i])$$
is the expected number of correct predictions that $\pi$ will make on $w$. They proved that $\text{Pred}_p(X)$ is related to the p-dimension of $X$ by
$$2(1 - \text{Pred}_p(X)) \leq \dim_p(X) \leq \mathcal{H}(\text{Pred}_p(X)) \tag{7.14}$$
(where $\mathcal{H}(\alpha)$ is the Shannon entropy of $\alpha$) and that these bounds are tight. If we call $\text{Pred}_p(X)$ the *upper feasible predictability* of $X$ and define the *lower feasible predictability* of $X$, $\text{pred}_p(X)$, in the same fashion, but with the limit superior in (7.13) replaced by a limit inferior, then we have the following dual of (7.14).

**Theorem 7.10.** *For all $X \subseteq \mathbf{C}$,*
$$2(1 - \text{pred}_p(X)) \leq \text{Dim}_p(X) \leq \mathcal{H}(\text{pred}_p(X)).$$

For each $s : \mathbb{N} \to \mathbb{N}$, let $\text{SIZE}(s(n))$ be the class of all (characteristic sequences of) languages $A \subseteq \{0,1\}^*$ such that, for each $n \in \mathbb{N}$, $A_{=n}$ is decided by a Boolean circuit consisting of at most $s(n)$ gates.

**Theorem 7.11.** *For each $\alpha \in [0,1]$, the class $X_\alpha = \text{SIZE}(\alpha \cdot \frac{2^n}{n})$ is pspace-regular, with $\dim_{\text{pspace}}(X_\alpha) = \text{Dim}_{\text{pspace}}(X_\alpha) = \dim(X_\alpha|\text{ESPACE}) = \text{Dim}(X_\alpha|\text{ESPACE}) = \alpha$.*

*Proof.* It was shown in [20] that $\dim_{\text{pspace}}(X_\alpha) = \dim(X_\alpha|\text{ESPACE}) = \alpha$. This proof also shows that the strong dimensions are $\alpha$. □

**Acknowledgment.** The third author thanks Dan Mauldin for extremely useful discussions.